\documentclass[aps, twocolumn, showpacs]{revtex4-1}
\usepackage{amsmath}
\begin{document}
\title {Real wave equations, spin origin of charge, and outlook on particle physics}
\author{S. C. Tiwari \\
Department of Physics, Institute of Science, Banaras Hindu University, Varanasi 221005, and \\ Institute of Natural Philosophy \\
Varanasi India\\}
\begin{abstract}
Radical revision on the conventional spacetime picture, illusion or emergent phenomenon, has been the focus of the speculations
on the unified frameworks for fundamental interactions in nature. Compared to strong experimental credence to the standard model of 
particle physics there is practically no relation of these speculations with the world of real particles. In this paper we present
a new conceptual framework for particle physics in which nontrivial geometry and topology of space and time have fundamental reality. 
To develop this model we proceed with the analysis of the standard wave equations and make them real using the transformation rule
$i=\sqrt{-1} \rightarrow C=\begin{bmatrix} 0 & 1 \\ -1 & 0 \end{bmatrix}$. Topological significance is 
attached to $i$ in terms of a point defect in 1D directed line. 
New interpretation of Schroedinger wave equation and $Z_2$ vortex throw light on the nature of spin. A new meaning is also
obtained on U(1) gauge symmetry and charge. Novel properties of C-matrix and known applications are presented. Topological
origin of spin is inferred from the interpretation of Pauli algebra that they signify 2D directed area in phase space as topological
obstruction. Electron magnetic moment decomposition in QED serves the basis for the proposition: spin origin of charge (SOC). A dynamical
logarithmic spiral geometric structure comprising of 2+1D braids, 3 vortices with associated three 2-spinors is envisaged based on SOC.
It is termed as meta-electron; meta-neutrino has 2-vortex structure. These are the only building blocks of all particles having the
knotted vortex structures in our model.
Coupling strengths of the three fundamental interactions are related with the magnitudes of the spin angular momentum of
constituent spinors of the meta-electron.
Unification has radically new paradigm compared to gauge theories: effective coupling constant with weight factors. To put the 
role of $i$ in perspective a discussion on previous works is also presented. Outlook on particle physics is elaborated in this new framework.
\end{abstract}
\pacs{12.90.+b}
\maketitle
\section{\bf Introduction}

The standard model in particle physics has enormous experimental support. However alternative ideas and extensions are being of great interest
to break the perceived impasse in the unification goal. 
We investigate real wave equations with a new approach in this paper: the main aim is to gain new insights on the nature 
of elementary particles and unification of their interactions.
Modern quantum field theories (QFTs) work in natural units $\hbar = c =1$ such that for length and time one has the dimension of $(mass)^{-1}$. In contrast,
physical interpretation of relativity and quantum mechanics crucially depends on the explicit presence of $c$ and $\hbar$ respectively.
Relativistic wave equation for a scalar, a vector, a spinor and a tensor field, let us denote them by $\mathfrak{U}$ (suppressing the indices) is given by
\begin{equation}
\partial^\mu \partial_\mu \mathfrak{U} = \nabla^2 \mathfrak{U} -\frac{1}{c^2} \frac{\partial^2 \mathfrak{U}}{\partial t^2} =0 
\end{equation}
Here the vacuum velocity of light $c$ is an integral part of the wave equation (1), however Planck constant (divided by $2 \pi$) $\hbar$ 
and imaginary unit $i = \sqrt{-1}$ do not appear in this equation. If quantum wave equation must necessarily contain $\hbar$ then
Eq.(1) is purely a classical relativistic wave equation with the assumption that the field $\mathfrak{U}$ has appropriate relativistic
Lorentz transformation property. In the literature, quite often the relativistic wave equation refers to quantum relativistic
wave equations, e. g. Dirac equation for the electron.

In the nonrelativistic Schroedinger wave equation  the distinct feature is the presence of both $(\hbar, i)$ that
gives it the mysterious characteristic \cite{1}, of course, the velocity of light has no role in this equation
that we write for free particle
\begin{equation}
-\frac{\hbar^2}{2 m} \nabla ^2 \Psi_s = i \hbar \frac{\partial \Psi_s}{\partial t} 
\end{equation}

Relativistic Schroedinger equation or Klein-Gordon (KG) equation is a scalar relativistic wave equation for a particle with non-zero rest mass
\begin{equation}
\nabla^2 \Phi - \frac{1}{c^2} \frac{\partial ^2 \Phi}{\partial t^2} = \frac{m^2 c^4}{\hbar^2} \Phi 
\end{equation}
Dirac equation for the electron is a first-order derivative wave equation for the Dirac spinor
\begin{equation}
i \hbar \frac{\partial \Psi_d}{\partial t} = - i \hbar c ~ {\bf \alpha}.{\bf \nabla} \Psi_d + \beta m c^2 \Psi_d 
\end{equation}
Comparing Eq.(3) and Eq.(4) we find that while $\hbar$ and $c$ are present in both equations, $i$ does not appear in Eq.(3), therefore,
the field $\Phi$ could be real or complex. The requirement that the Dirac wavefunction $\Psi_d$ satisfies the second-order derivative
relativistic wave equation of the form (3) leads to the condition that the quantities $\alpha_x, \alpha_y, \alpha_z, \beta$ anticommute
in pairs and the square of each is unity. It can be shown that ${\bf \alpha},~{\bf \beta}$ are $4 \times 4$ Hermitian matrices, and
$\Psi_d$ is a four component complex wavefunction that can be represented as a column matrix with four rows.

A nice exposition on the mathematical and physical foundations of the wave equations is given by Corson \cite{2}. Of special interest is a heuristic but
illuminating discussion on spinors, geometry, and spinor wave equations in \cite{2}. Mathematical concept of a spinor in three-dimensional
space is originally due to Cartan \cite{3}. The structure of the wave equations has inspired numerous new developments; the most striking is the 
discovery of the Dirac equation adhering to fundamental physical principles of quantum mechanics.

The set of equations (1)-(4) has well-known applications in physics. However the intrigue on the physical interpretation of the Schroedinger
equation (2) has never ended; even after the advent of quantum information science new questions on the foundations of quantum mechanics
have arisen. Regarding the Dirac equation (4), the meaning of zitterbewegung and its relation with spin have inspired a vast literature.
In the present paper the significance of the Planck constant $\hbar$, the relativistic-invariant mass parameter $m$, and the imaginary unit $i$
is examined in a new approach. The importance of the limit $\hbar \rightarrow 0$ for the quantum to classical transition has varied perception
among physicists. Here we focus on the limit $m \rightarrow 0$. Note that massless fields and particles have many attractive mathematical
and geometrical properties. One of the surprising consequences of massless limit, not appreciated in the literature, is that the Plamck
constant disappears from the wave equations in general. The imaginary unit also disappears in the Schroedinger equation, however its
presence in the Pauli matrix ensures that the Dirac equation cannot be made real just by taking the massless limit.

Recall that the complex numbers have enriched the mathematical analysis, and they have profound geometrical and topological properties. 
Complex representation of real physical quantities proves to be a powerful and convenient calculational tool. The presence of $i$ in the 
wave equations at a fundamental level, for example, in the Schroedinger and Dirac equations, however implies that the wavefunctions have 
to be necessarily complex. In fact, the initial apparent embarassment has been turned to a virtue relating complex wavefunctions with
the electrically charged particles. Moreover, the concept of anti-particles and spin-half interpretation of the Dirac equation nicely
co-exist with the complex wave equation. Majorana in 1937 was able to obtain real version of the Dirac equation \cite{2}. Majorana's
speculations did not find favor at that time, however it has found strong contemporary relevance, for example, in the form of Majorana
neutrinos, supersymmetric partners like photino for photon, and Majorana modes in condensed matter systems. The search for real
wave equations has another, more cogent justification: the only wave equation founded on direct experimental laws, namely the Maxwell equations.
is real.

The discussion in the next section brings out the salient aspects on the structure of wave equations throwing light
on the role of fundamental constant $\hbar$.
The critical review leads to the following question. Is there any relation between the imaginary unit and spin? A new
approach is articulated based on the equivalence between $i$ and a matrix C introduced in \cite{3}
\begin{equation}
 C = \begin{bmatrix} 0 & 1 \\ -1 & 0 \end{bmatrix}
\end{equation}
that has the property that $C^2 = -1$; in fact, the unity here is a $2 \times 2$ matrix that we denote by $I_2$. We propose a transformation
\begin{equation}
i \rightarrow C
\end{equation}
as a technique to obtain real wave equations. A new perspective on geometry and topology of complex
numbers and the properties of the matrix $C$ constitute section III.
An important consequence on the nature of topological defect is used to analyze $Z_2$ vortex for a real Hermitian Hamiltonian describing 
a two-level quantum system \cite{4}. It is shown in section IV
that the nonrelativistic Schroedinger equation (2) under the transformation (6) becomes a pair of coupled wave equations with real wavefunctions. 
This transformation is not a trivial re-statement of the Schroedinger equation: new physics to interpret spin and imaginary unit arises here. Note
that the unconventional connection of spin with Schroedinger equation was first pointed out by Gurtler and Hestenes \cite{5} based on a
different reasoning. In section V the transformation (6) is applied to 2-spinor Weyl equation and Dirac equation to obtain real wave equations; 
relationship of new form of real Dirac equation with Majorana equation is discussed. Intriguing role of Pauli matrix $\sigma_2$ in real form
is discussed and the Pauli algebra is given a topological interpretation in terms of the directed area defect in phase space. A radically new
perspective emerges in our approach for particle physics presented in section VI. In section VII the present work on the interpretation of $i$ is
discussed in the context of the previous literature. Concluding remarks constitute the last section.

\section{\bf The Nature of Wave equations}

Physical arguments to examine new hypotheses and mathematical formalisms are essentially based on three principal ingredients: the observed
physical phenomena, empirical facts/data, and physical principles. All the three ingredients are not static and fixed; they have evolved into new forms
with the advances in experiments and theory. Past discards may re-emerge with new physical relevance. It is crucial to understand this 
continuously changing conception of physical reality; here we focus on wave equations in this conceptual framework.

Foremost example is that of the discovery of the Dirac equation (4). Dirac found KG equation (3) unsatisfactory on physical grounds: the 
probability density not being positive-definite, and admissibility of negative energy solutions. Probability interpretation for 
Dirac current is correct but the issue of negative energy persists. Natural occurrence of spin and electron spin magnetic
moment in the Dirac equation, and prediction of anti-particles constitute landmark achievement of the Dirac equation with the discovery
of positron by Anderson in 1932.

Is KG equation unphysical? Yukawa's field for the interaction between neutron and proton \cite{6} was described by the equation of the 
form (3). Now we know that KG equation describes spinless neutral as well 
as charged particles for real and complex wavefunctions respectively. Much later Dirac proposed a positive energy relativistic
wave equation \cite{7} that gives integral spin values. In a subsequent paper \cite{8} its connection with Majorana equation was pointed out.
Majorana in 1932 obtained an equation, see \cite{2}, having unusual properties: infinity of mass values with the largest mass having zero 
spin that increases progressively to infinity as mass tends to zero. This contradicts experimental observations. Dirac, restricting his
equation to only one mass value, found spin to depend on the momentum. He traces this unsatisfactory result to 'an obscurity in the definition of spin'.

Is there any mass-spin relationship? Today there exist hundreds of observed elementary particles (hadrons and leptons), and gauge bosons (photon,
weak gauge bosons and gluons) showing no connection between mass and spin. Therefore let us try to understand preceding remark of Dirac that has
a specific context in his pulsating spherical shell model. In the textbooks \cite{9} following Dirac, spin angular momentum (SAM) is usually 
interpreted as a term added to the orbital angular momentum (OAM) to obtain the constant of motion in a central field. To define OAM 
${\bf r} \times {\bf p}$ one needs a set of space coordinates. Dirac argues that for a particle at rest the coordinate choice does not matter, 
and SAM is well-defined. A gauge-invariant set of coordinates is shown by him to avoid the problem in defining SAM for non-zero momentum \cite{8}.
There does exist a wave equation, namely, the Majorana equation \cite{10} in which infinite number of spin values $s$ determine the mass eigenvalues
$M/(s +\frac{1}{2})$, where $M$ is arbitrary constant mass parameter. The kind of mass-spin relation found by Dirac \cite{7, 8} and 
Majorana \cite{10} at present has no experimental support, and represents a hypothetical curiosity.

However there does exist a useful classification in which non-zero mass and zero mass particles and their spin have distinct representations,
see Chapter IV in \cite{2}. Unlike Dirac method, here one has general considerations based on the Casimir invariants
\begin{equation}
P= p_\mu p^\mu 
\end{equation}
\begin{equation}
 W = W_\mu W^\mu
\end{equation}
Here the Pauli-Lubanski pseudo-vector is $W_\mu = - \frac{1}{2} \epsilon_{\mu \nu \lambda \sigma} J^{\nu \lambda} p^\sigma$. To define SAM for 
a non-zero mass particle one goes to the rest frame in which OAM is zero, and wavefunction depends only on time. Irreducible representation 
under spatial rotation gives $2 s+1$ independent components for spin $s$ that may have half-integral or integral values, and Lorentz
invariance extends the validity to arbitrary reference frame. For zero mass there is no rest frame. One specifies a reference frame in which
momentum is directed along a particular axis; OAM is zero along this axis and any angular momentum along this axis has to be SAM. In this case
there are only two independent components as compared to $2 s +1$ for non-zero mass particles.

Wave equations irreducible under the inhomogeneous  Lorentz group (or Poincare group) are discussed in the last section 40 in the monograph
\cite{2}. Three general classes are presented: 1) discrete spin, non-zero rest mass, 2) discrete spin and zero mass, and 3) continuous 
spin and zero mass. The third one has apparently no physical realization. Corson explains that infinitely many states of polarization are 
described by a continuous spin variable, and half-integral (integral)  has meaning in terms of double-valued (single-valued) representation.
The scalar wavefunction in coordinate representation satisfies the following equations
\begin{equation}
\partial_\mu \partial^\mu \Phi =0 
\end{equation}
\begin{equation}
\xi ^\mu \partial_\mu \Phi =0
\end{equation}
\begin{equation}
(\xi^\mu \xi_\mu + \lambda^2) \Phi =0 
\end{equation}
\begin{equation}
(\frac{\partial^2}{\partial \xi^\mu \partial x_\mu} - \Sigma ) \Phi =0 
\end{equation}
Here auxiliary variable $\xi^\mu$ is a space-like 4-vector of length $\lambda$ orthogonal to $p_\mu$, and $\Sigma$ is a real positive number. 
It can be proved that though the wavefunction cannot be localized the norm is Lorentz invariant and positive-definite. The wavefunction
for half-integral representation can be constructed if additional 4-valued spin variable is introduced.

Zero rest mass fields (or particles) have great interest in mathematical physics, for example, zero-length is invariant in Weyl space, 
wave equations have conformal invariance, and in sheaf cohomology and twistors. Penrose \cite{11} shows that the solution of spin $s=0, \frac{1}{2}, 1 ...$
zero mass free field equations can be obtained from a contour integral of an arbitrary analytic function of three complex variables; this
paper of Penrose treats the problem independently of twistor theory. Physical reality, however seems to indicate only one zero mass field, namely
the photon. Massless gluons cannot be observed free due to color confinement and experiments show that neutrinos are not massless \cite{12}. Surprisingly
massless quasi-particles, including Dirac fermions, are finding many applications in condensed matter systems. Do abstract mathematical objects
of third category have any role at a fundamental level in physics?

Let us consider the wave equations afresh. Interpretation of Schroedinger equation has been sought from various angles in the literature \cite{1}.
A new line of thinking is to seek the limit $m \rightarrow 0$ \cite{13}. Eq.(2) reduces to the Laplace equation, and for relativistic invariance
it is generalized to the usual wave equation of the form (1). Obviously $\hbar$ and $i$ do not appear in this wave equation.

Dirac equation (4) in the massless limit assumes the form
\begin{equation}
- c ~ {\bf \alpha}. {\bf \nabla} \Psi_d = \frac{\partial \Psi_d}{\partial t} 
\end{equation}
Note that $i, \hbar, \beta$ disappear in Eq.(13). Anticommutation relations for ${\bf \alpha}$ matrices can be satisfied by $2 \times 2$
Pauli matrices ${\bf \sigma}$
\begin{equation}
 \sigma_1 =\begin{bmatrix} 0 & 1 \\ 1 & 0 \end{bmatrix}; ~ \sigma_2 =\begin{bmatrix} 0 & -i \\ i & 0 \end{bmatrix}; ~ 
 \sigma_3 =\begin{bmatrix} 1 & 0 \\ 0 & -1 \end{bmatrix}
\end{equation}
Instead of 4-spinor Dirac wavefunction we now have 2-spinor wavefunction in the following equation
\begin{equation}
 - c {\bf \sigma}.{\bf \nabla} \Psi_w = \frac{\partial \Psi_w}{\partial t}
\end{equation}
Due to $i$ in Pauli matrices the wavefunction $\Psi_w$ has to be complex. The usual SAM operator is defined to be
${\bf S} = \frac{1}{2} \hbar {\bf \sigma}$. However in the absence of $\hbar$ in Eq.(15) one is free to introduce arbitrary angular momentum unit; one
can set a continuous dimensionless variable $y$ in the re-defintion
\begin{equation}
 \tilde{\bf S} = \frac{1}{2} y \hbar {\bf \sigma}
\end{equation}
The spinor wavefunction is double-valued but SAM can be continuous. The classification of massless particle and continuous spin acquires a new
physical interpretation.

This marks a radical departure from the traditional view on spin. Does this idea contradict physical principles? 
To understand this question let us examine how the concept of spin has
evolved. Quantized angular momentum in the old quantum theory at first appeared too radical to the physicists. Later Pauli described spin to be a non-classical
intrinsic characteristic having no classical picture. The enigma of spin continued even after Dirac equation and Stern-Gerlach experiment \cite{14}.
Now we have a sound quantum field theory
\cite{2, 12}, spin-statistics relation and neat classification of elementary particles as bosons and fermions. The scope of the standard physics has been
enlarged based on speculations on fractional spin, anyons, and supersymmetry (SUSY). SUSY plays central role in supergravity and superstrings 
\cite{15}, and models beyond the Standard Model \cite{12}. Note that SUSY particles continue to remain elusive even at TeV energy scales at LHC, CERN.
The point is that so far as the physical principles are concerned in the present theoretical scenario, the new 
idea is not in conflict with them. The only objection could arise if one believes in the traditional picture.

Note that the spinor wavefunction $\Psi_w$ also satisfies the massless second-order wave equation. Now a topological defect in a Euclidean 
plane $R^2 -\{0\}$, i. e. the origin is removed, is realized for a singular vortex.
For example, the scalar wave equation of type (1) can be solved assuming following form
\begin{equation}
\Phi = \Phi_0(x, y) e^{i(k z - \omega t)} 
\end{equation}
where $\omega^2 = k^2 c^2$, and $\Phi_0$ satisfies Laplace equation in 2-dimension. In a circular coordinate system $(r, \theta)$ the singular 
solution corresponds to $\Phi_0 = \beta \theta$, where $\beta$ is an integration constant. Here ${\bf \nabla} \Phi$ is singular at $r=0$.
The vortex for the spinor would be a propagating massless one, therefore, we need a mechanism to explain the origin of mass.
In the literature, another approach has been used obtaining localized non-spreading (soliton-like) solutions of the massless wave equation to
represent massive particles. In the present work mass is not an intrinsic property at a fundamental level, therefore we seek a different mechanism.

For this purpose, let us analyze the role of mass in Dirac equation written in the Lorentz covariant form
\begin{equation}
(i \hbar \gamma^\mu \partial_\mu - mc ) \Psi_d =0 
\end{equation}
where the gamma matrices satisfy the anti-commutation relations or Dirac algebra. Dirac 4-spinor has 2-dimensional representation, and 
may be  written as
\begin{equation}
\Psi_d =\begin{bmatrix} \Psi_L \\ \Psi_R \end{bmatrix}
\end{equation}
One observes that mass couples $\Psi_L$ and $\Psi_R$; if $m=0$ we get de-coupled Weyl equations. A physical interpretation of this coupling 
has to be searched.  An interesting formal derivation of Dirac equation \cite{16} is
worth mentioning in this connection: electron travels at the speed of light, just like $\Psi_L$ or $\Psi_R$, and flips chirality at 
random times with the rate of flips related with mass. Thus stochastic origin of mass is an attractive idea. The presence of $i$ makes it
necessary to invoke analytic continuation in the stochastic approach.  It would be of value to have real wave equations.

In the wave equations one simply assumes that the description of charged particles requires complex wavefunctions. The
probability density and the probability current density get charge density and current density interpretation putting $e$ by hand as
a multiplying factor. Putting by hand means the implied arbitrariness, for example, one multiplies  $\Psi^*_s \Psi_s$ by $m$ 
to interpret it as mass density. In
the Dirac current $\bar{\Psi_d} \gamma^\mu \Psi_d$ multiplication by $e$ gives the charge current density 4-vector. In QED the calculation
of renormalized mass and charge is carried out once bare mass and charge are postulated. In the modern version, QED is a U(1) gauge
field theory, and the Dirac current is a Noether current corresponding to U(1) symmetry. What is this internal U(1) space? One has
such internal space for lepton number too. In the Standard Model \cite{12} the internal spaces are assumed for various guage symmetries. 
Internal space seems more an artefact compared to the intuitive concept of physical space and time. Preceding discussion suggests that
$i$ may be related with spin, and logically charge has some kind of spin (fractional!) interpretation.

\section{\bf Geometry and Topology}

Cartan introduced the matrix $C$ to define a conjugate spinor or a spinor of second type  \cite{3}. The proposition (6) 
to obtain real wave equations, to the author's knowledge, is being used for the first time following a recent preliminary report \cite{17}.
Survey of the literature, however, shows at least one example where this transformation has been used \cite{18}. In the famous textbook 
on superstrings \cite{18} the authors have string-theoretic motivation to describe a vacuum state in a 10-dimensional space-time of the form
$T^4 \times K$ where $T^4$ is, or may be, a Minkowsian space-time and $K$ is a compact 6-manifold. Seeking SU(3) holonomy the embedding
of SU(3) in SO(6) is demonstrated using the replacement (6) transforming a complex number $a + i b$ as follows
\begin{equation}
a + i b ~ \rightarrow ~ a I_2 + b C 
\end{equation}
A complex $3 \times 3$ unitary matrix becomes a $6 \times 6$ orthogonal real matrix. In this
section we put the transformation rule (6) on a more secure foundation, and show that it is not a mere re-statement but offers new insights,
specially on the nature of topology.

{\bf III-A. ~ Complex Variables}

Relevant properties of complex numbers and their functions given in the textbooks are summarized here for the self-contained
discussion. Any complex number $z= x + i y$ can be represented geometrically on Argand diagram: a 2-dimensional $(x, y)$ plane in
Cartesian coordinate system or in polar coordinates $(r, \theta)$, where $r$ is the modulus $|z|$ and angle $\theta$ is the argument of $z$.
Note that $arg~ z$ is not unique; the principal value of  $arg ~ z$ is defined by $-\pi < arg ~z \leq \pi$. 
In analogy to the real analysis, we define the set of all points $z$ such that
$|z - z_0| < \epsilon$ as a neighborhood of a point $z_0$, where $\epsilon$ is a positive number. Complex function $w(z) = u + i v$
of the complex variable $z$ defined equivalently as $w(z) = u(x,y) +i v(x, y)$ is continuous at $z_0$ if for a given $\epsilon > 0$ we have
a number $\delta$ such that $ |w(z) -w(z_0)| < \epsilon$, for all points $z$ satisfying $|z - z_0| < \delta$.

A function $w(z)$ that is single-valued and differentiable at every point of a domain $D$ is said to be regular in the domain $D$. 
Necessary and sufficient conditions for a function to be regular are that the Cauchy-Riemann equations are satisfied. There exist interesting
multi-valued functions, for example,
\begin{equation}
 w(z) = \sqrt{z}
\end{equation}
is not single-valued. Substituting $z=r e^{i \theta}$ in (21) shows that i) for fixed $\theta$, $w_1 = |\sqrt{r}| e^{i \theta/2} $ and
$w_2 = -|\sqrt{r}| e^{i \theta/2} $ are the only continuous solutions, ii) varying $\theta$ from $0$ to $2 \pi$, the variable $z$ describes a 
circle of radius $r$ about the origin, and $w_1$ varies continuously becoming discontinuous at $\theta = 2 \pi$ becoming $w_2$, and iii) if $z$
traces the circle second time $w_1$ returns to itself. Thus the function (21) is not single-valued continuous on the whole complex plane. 
Geometrically the two-valuedness can be represented in terms of two branches or Riemann sheets on which it is single-valued, i. e. two 
complex planes and a cut from origin to infinity along the positive real axis. It is important to remember that the cut-line is not unique, 
however the point $z=0$ is unique, and it is termed the branch point. Thus nontrivial topology and geometry are significant in complex
functions/analysis.

{\bf III-B. ~ C Matrix}

Cartan \cite{3} considers a 3D Euclidean space and postulates an isotropic, i. e. zero-length vector $(A_1, A_2, A_3)$
\begin{equation}
A_1^2 +A_2^2 +A_3^2 =0 
\end{equation}
Spinor is defined as a pair of quantities $\xi_1, \xi_2$ such that
\begin{equation}
 \xi_1 = \pm \sqrt{(A_1 - i A_2)/2}
\end{equation}
\begin{equation}
  \xi_2 = \pm \sqrt{-(A_1 + i A_2)/2}
\end{equation}
He calls spinor as a kind of directed or polarized isotropic vector. A vector ${\bf X}$ has a matrix associated with it
\begin{equation}
X = \begin{bmatrix} x_3 & x_1 - i x_2 \\ x_1 + i x_2 & -x_3 \end{bmatrix}
\end{equation}
That the Pauli matrices (14) are the matrices associated with the basis vectors follows from (25). Cartan introduces the matrix $C$  and 
presents its important properties. Besides $C^2 = -I_2$, following relations are also given
\begin{equation}
 C^T = -C; ~ C C^T = I_2
\end{equation}
Action of $C$ on the vector ${\bf X}$ gives
\begin{equation}
 C X = -X^T C
\end{equation}
A spinor conjugate to $\xi$ is defined as
\begin{equation}
 \xi_c = i C \bar{\xi}
\end{equation}
The conjugation operation is not involution since under conjugation $\xi_c \rightarrow - \xi$. The conjugate spinor is termed 
a spinor of the second type.

In a pseudo-Euclidean space, the matrix associated with a vector is real
\begin{equation}
 X = \begin{bmatrix} x_3 & x_1 -  x_2 \\ x_1 +  x_2 & -x_3 \end{bmatrix}
\end{equation}
In this case an isotropic vector is associated with real component spinors.

Remarkably, in the spinor analysis of van der Waerden \cite{19} the matrix $C$ denoted by Levi-Civita symbol, $\epsilon_{ij}$, 
plays an important role in relating covariant and contravariant spinors.
\begin{equation}
 C^{-1} = C^T = \epsilon^{ij} = \begin{bmatrix} 0 & -1 \\ 1 & 0 \end{bmatrix}
\end{equation}
\begin{equation}
 C= \epsilon_{ij} =  \begin{bmatrix} 0 & 1 \\ -1 & 0 \end{bmatrix}
\end{equation}

We have found some novel aspects related with the matrix $C$ that we present here. First, it follows from (29) that the matrices
associated with the basis vectors in the pseudo-Euclidean space are real
\begin{equation}
 \sigma_1 ;~ \sigma_3 ; ~C^T
\end{equation}
Elementary matrix algebra establishes that the three matrices (32) anti-commute
\begin{equation}
 \{ \sigma_1 , \sigma_3 \} = \{ \sigma_3 , C^T \} = \{ \sigma_1 , C^T \} = 0
\end{equation}
and their commutators are
\begin{equation}
 [ \sigma_1 , C^T ] = 2 \sigma_3
\end{equation}
\begin{equation}
 [ \sigma_1 , \sigma_3 ] =2 C^T
\end{equation}
\begin{equation}
 [ C^T , \sigma_3 ] = 2 \sigma_1
\end{equation}
Expressions (30) and (31) show that if we assume the replacement rule (6) then we must have $- i \rightarrow C^T$. It may appear puzzling at 
first that the Pauli matrix $\sigma_2$ is equivalent to the identity matrix
\begin{equation}
 \sigma_2 = C C^T = I_2
\end{equation}
However expression (37) merely represents the factorization of unity in the form $1 = - i^2 = (i) (-i)$. The standard
matrix $\sigma_2$ is a mixed representation combining real matrix $C^T$ with the imaginary unit $i$.

The set $(I_2, \sigma_i)$ forms a finite non-abelian group under matrix multiplication since $\sigma_i^{-1} = \sigma_i$. The set $(I_2, \sigma_3)$
forms its abelian subgroup. On the other hand the set of real matrices $(\pm I_2, \sigma_1, \sigma_3, C, C^T)$ forms group under matrix multiplication, 
and $(\pm I_2, C , C^T)$ is an abelian subgroup.

Antisymmetric matrix $C~(C^T)$ has a number of applications. Let us discuss the Lie algebra: the generators of $SU(2)$ group in the defining representation 
are $\frac{\sigma_i}{2}$. In the case of group $SL(2,Z)$
the group generators are
\begin{equation}
 C^T;~ T = \begin{bmatrix} 0 & -1 \\ 1 & 1 \end{bmatrix}
\end{equation}
where $T^3 = -I_2$. Note that $SL(2, Z)$ is a subgroup of $SL(2, R)$ which has the real matrix
\begin{equation}
 A = \begin{bmatrix} a & b \\ c & d \end{bmatrix}; ~ ad -bc =1
\end{equation}
In $SL(2, Z)$, numbers $(a,b,c,d)$ are integers, and could be generated by the product of finitely many factors of the generators (38).

In physics applications $SL(2,Z)$ appears in the electric-magnetic duality and Schwinger quantization of electric-magnetic charges. In
supersymmetric gauge theory the vacuum angle $\theta$ combined with charge $e$ gives a complex parameter
\begin{equation}
 S= \frac{\theta}{2 \pi} + i \frac{4 \pi}{e^2}
\end{equation}
Here magnetic charge is $Q_m = \frac{n}{e}$ and electric charge is $Q_e = e (m + \frac{n \theta}{2 \pi})$, where $n,m$ are integers; the 
duality has the group $SL(2,Z)$. In nonlinear sigma-models an important role is played by a non-compact target space, namely 2D Poincare plane.
The isometry group of the Poincare plane is $SL(2,R)$. The matrix $C$ is sometimes called symplectic metric, and 
symplectic group and manifolds are useful in duality and supersymmetric theories. If $S$ is a real $2\times 2$ matrix then symplectic group Sp(2)
is defined by $S^T C S=C$. For details we refer to an excellent book on supergravity \cite{20}.

Finally we explore topological aspects. Abelian unitary group U(1) has just the phase factor
\begin{equation}
 U(1) = e^{i \theta} = cos \theta + i sin \theta
\end{equation}
The transformation rule (6) in (41) gives 2D rotation group 
\begin{equation}
 SO(2) = cos \theta I_2 + sin \theta C = \begin{bmatrix} cos \theta & sin \theta \\ -sin \theta & cos \theta \end{bmatrix}
\end{equation}
This well-known result is obtained here in a simple way using $C$. Note that (41) represents a circle of unit radius in the Argand diagram,
whereas (42) corresponds to the rotation in real 2D plane. Let us re-visit real analysis. A real number $a$ is positive or negative according
as $a>0$ or $a<0$; i. e. on $(R^+ , R^-)$. Geometrically, specifying a point $O$ on a straight line divides 
the line: conventionally right is positive and left is negative.
Such a directed line has a unique  point on the line corresponding to every real number, and to every point on the line corresponds
a unique real number; sometimes it is phrased as Dedekind-Cantor axiom. There is no continuous transformation that maps points on $R^+$ 
to points on $R^-$, and one defines right and left hand limits as $0^+$ and $0^-$. Introducing imaginary axis it is possible to have a 
continuous transformation on the complex plane, using (41) to map points on $(R^+; R^-)$. An alternative is to make $O$ as a 
unique point representing discontinuity, and use $C$ matrix to transform $R^+ \rightarrow R^-$. A real number gets transformed as follows
\begin{equation}
 \begin{bmatrix} 0 & 1 \\ -1 & 0 \end{bmatrix} \begin{bmatrix} a \\ 0 \end{bmatrix} = \begin{bmatrix} 0 \\ -a \end{bmatrix}
\end{equation}
Instead of (43) we may assume that $C$ is used only once for $0^+$
\begin{equation}
\begin{bmatrix} 0 & 1 \\ -1 & 0 \end{bmatrix} \begin{bmatrix} 0^+ \\ 0 \end{bmatrix} = \begin{bmatrix} 0 \\ -0^+ \end{bmatrix} 
\end{equation}
Two segments of the directed line are joined by a discontinuous jump $0^+ - 0^- = \epsilon_0$, where $\epsilon_0$ is an infinitesimal real number.
Both segments now have a continuous transformation, however the point $O$ analogous to the branch point for a multi-valued complex 
function discussed in section III-A behaves as a point topological defect on 1D directed real line.

{\bf III-C. $~ Z_2$ Vortex}

Wilczek \cite{4} considers a simple quantum problem of level-crossing in a two-level system described by a real Hermitian Hamiltonian
\begin{equation}
 H(x, y) = \begin{bmatrix} x & y \\ y & -x \end{bmatrix}
\end{equation}
with energy eigenvalues
\begin{equation}
 E = \pm \sqrt{(x^2 + y^2)}
\end{equation}
The geometry of level crossings is discussed calculating the wavefunctions. The positive energy eigenfunction is 
\begin{equation}
 \Psi_+(x,y) = e^{- i \sigma_2 \frac{\phi}{2}} \begin{bmatrix} 1 \\ 0 \end{bmatrix}
\end{equation}
\begin{equation}
 \Psi_+(x,y) = \begin{bmatrix} cos \frac{\phi}{2} \\ sin \frac{\phi}{2} \end{bmatrix}
\end{equation}
Here $\phi = tan^{-1} \frac{y}{x}$. The sign of $\Psi_+$ is reversed as $\phi$ changes from $0 ~to ~2 \pi$; this discrete topology
is termed a $Z_2$ vortex. The interesting question is raised that for a complex energy eigenfunction
\begin{equation}
 \tilde{\Psi}_+ (x,y) = e^{\frac{i \phi}{2}} \begin{bmatrix} cos \frac{\phi}{2} \\ sin \frac{\phi}{2} \end{bmatrix}
\end{equation}
$\phi \rightarrow \phi + 2 \pi$ leaves invariant (49) and the discrete topology of (48) disappears: why? The answer 
is quite illuminating. Aharonov-Bohm like geometric phase arises with the gauge potential
\begin{equation}
 A_\phi = -\frac{i}{2}
\end{equation}
\begin{equation}
 e^{\int A_\phi ~d\phi} = e^{-\frac{i\phi}{2}}
\end{equation}
and restores the sign-reversal found for real wavefunction (48). Thus the existence of a $Z_2$ vortex for the nondegenerate
level crossings of a real Hamiltonian is independent of the formulations: a smooth and continuous wavefunction in two patches
on the circle with a transition factor, or smooth continuous function over the whole circle with a globally nontrivial
Aharonov-Bohm like gauge potential.

Let us re-examine Wilczek's analysis using $C$ matrix. It is straightforward to verify that replacing $-i \rightarrow C^T$, and 
$\sigma_2 \rightarrow I_2$ the exponential pre-factor in (47) becomes $e^{C^T \frac{\phi}{2}}$, and the eigenfunction (48) is 
immediately obtained. The Hamiltonian (45) has been re-written in \cite{4} in the following form
\begin{equation}
 H = \sqrt{(x^2 +y^2)} e^{- i \sigma_2 \frac{\phi}{2}} \sigma_3 e^{i \sigma_2 \frac{\phi}{2}}
\end{equation}
The complex eigenfunction $\tilde{\Psi}_+$ given by (49) replacing $i \rightarrow C$ is nothing but the eigenfunction of $\sigma_3$
\begin{equation}
 \tilde{\Psi}_+ \rightarrow \begin{bmatrix} 1 \\ 0 \end{bmatrix}
\end{equation}
It is not surprising because the transformation of the wavefunction (49) has to accompany a transformation of the Hamiltonian (52) resulting
into $\sigma_3$. The geometric phase factor (51) in the expression (53) restores the real eigenfunction (48); here it is not a phase
factor but real exponential $e^{C^T \frac{\phi}{2}}$. The important conclusion from our analysis based on the real wavefunctions using 
the rule (6) is that the discrete topology $Z_2$ has origin in the discontinuity (nontrivial transition factor) in either formulation 
discussed by Wilczek. 

\section{\bf Spin in the real Schroedinger equation}

The imaginary unit $i$ in the Schroedinger equation, for example, for a free particle Eq.(2), has been a source of mysterious interpretations
in the literature, see \cite{1}, and even today the meaning of the physical reality of the wavefunction $\Psi_s$ remains unsettled. If
one could transform Schroedinger equation to a real wave equation then a new pathway for resolving foundational questions may be envisaged.
A recent note \cite{17} makes an attempt in this direction emphasizing the relation with the stochastic interpretation \cite{21}. Here we focus
on the question whether $i$ in the Schroedinger equation hides spin in some way. Gurtler and Hestenes \cite{5} examine the consistency
between Dirac, Pauli, and Schroedinger equations, and arrive at an unorthodox result: Schroedinger equation describes a particle in a spin
eigenstate not a spinless particle. Briefly stated, the argument is simple and logical based on the theory reduction: Pauli equation reduces
to the Schroedinger equation when the magnetic field is negligible (or zero), but the wavefunction now represents spin eigenstate
\begin{equation}
 \Psi_s = \begin{bmatrix} \Psi_s^0 \\ 0 \end{bmatrix}
\end{equation}
As a consequence the magnetization current is nonzero. For $\sigma_3$ diagonal the calculated magnetization for the wavefunction (54) is
\begin{equation}
 m_3 = \frac{e \hbar}{2 m c} \Psi_s^\dagger \sigma_3 \Psi_s = \frac{e \hbar}{2 m c} \rho_s
\end{equation}
and there is a non-vanishing magnetization current ${\bf \nabla} \times {\bf m}$. Therefore the usual Schroedinger  charge current density
is inconsistent with it. Unfortunately, as remarked by the authors, direct experimental proof for the spin state of the Schroedinger
particle is not possible since the detection of the magnetization current requires the magnetic field but then one makes use of 
the Pauli theory.

The important point is that the authors link the presence of complex numbers in the Schroedinger theory with the spin. Instead of theory
reduction followed in \cite{5}, i. e. Dirac to Pauli to Schroedinger theories, we propose a new approach based on the real Schroedinger
wave equation. Note that the present approach is fundamentally different than the one based on separating real and imaginary parts as
is done in de Broglie-Bohm theory. We employ the rule (6) that demands the wavefunction $\Psi_s$ to be a real spinor
\begin{equation}
 \Psi_s ~ \rightarrow ~\begin{bmatrix} \eta \\ \chi \end{bmatrix}
\end{equation}
Schroedinger equation (2) is transformed to coupled wave equations
\begin{equation}
 -\frac{\hbar^2}{2m} \nabla^2 ~  \eta = \hbar \frac{\partial \chi}{\partial t} 
\end{equation}
\begin{equation}
 -\frac{\hbar^2}{2m} \nabla^2 ~ \chi = -\hbar \frac{\partial \eta}{\partial t}
\end{equation}
Eqs. (57) and (58) in spinor form read
\begin{equation}
 -\frac{\hbar^2}{2m} \nabla^2 ~  I_2 \begin{bmatrix} \eta \\ \chi \end{bmatrix} = \hbar \frac{\partial }{\partial t} C
 \begin{bmatrix} \eta \\ \chi \end{bmatrix} 
\end{equation}
It is straighforward to derive the current continuity equation multiplying (57) by $\chi$ and (58) by $\eta$ and subtracting the 
resulting equations. We obtain
\begin{equation}
{\bf \nabla}.\tilde{\bf J} +\frac{\partial \tilde{\rho}}{\partial t} =0 
\end{equation}
\begin{equation}
\tilde{\rho} = \frac{(\eta^2 +\chi^2)}{2} 
\end{equation}
\begin{equation}
\tilde{\bf J} = \frac{\hbar}{2 m} ( \chi {\bf \nabla} \eta - \eta {\bf \nabla} \chi)
\end{equation}
From expression (62) one may introduce a current velocity field $\tilde{\bf v}$
\begin{equation}
\tilde{\bf v} = \frac{\tilde{\bf J}}{\tilde{\rho}} 
\end{equation}

Spinor formulation of the Schroedinger equation is a radical departure from the conventional approaches seeking analogy to hydrodynamics or
diffusion equation. Note that the standard Schroedinger equation (2) is a trivial special case when $\eta \propto \chi$; consistency
between (57) and (58) shows that the proportionality constant is $\pm i$, see \cite{17}.

How do we interpret spin? A simple calculation gives
\begin{equation}
 \begin{bmatrix} \eta & \chi \end{bmatrix} \sigma_3 \begin{bmatrix} \eta \\ \chi \end{bmatrix} = \eta^2 - \chi^2 = \tilde{S}
\end{equation}
Logically the quantity $\tilde{S}$ may be expected to have a relation with spin. To understand it let us return to the set of Eqs. (57) and (58);
multiplying them by $\chi$ and $\eta$ respectively and adding the reulting equations we obtain
\begin{equation}
 \frac{\hbar}{2} \frac{\partial \tilde{S}}{\partial t} = \frac{\hbar^2}{2 m} (\eta \nabla^2 \chi + \chi \nabla^2 \eta)
\end{equation}
The expression on the right hand side of (65) could be re-written as
\begin{equation}
 \frac{\hbar^2}{2 m} [ {\bf \nabla}. {\bf \nabla} (\eta \chi) - 2 {\bf \nabla} \chi . {\bf \nabla} \eta ]
\end{equation}
The curl of the current density (62) is calculated to be
\begin{equation}
 {\bf \nabla} \times \tilde{\bf J} = \frac{\hbar}{2 m} 2 ({\bf \nabla} \chi \times {\bf \nabla} \eta)
\end{equation}
In a special case when ${\bf \nabla} \chi$ and ${\bf \nabla} \eta$ are orthogonal, the second term in the expression (66) vanishes
and it becomes a total divergence. Integrating Eq.(65) over a volume $V$, transforming the divergence term to a surface integral
that is assumed to vanish at infinity, we finally arrive at the following important result
\begin{equation}
 \frac{\hbar}{2} \frac{\partial \int \tilde{S} dV}{\partial t} =0
\end{equation}
Equation (68) is suggested to have the physical interpretation that $\tilde{S}$ signifies the spin angular momentum density. It
is worth making two important remarks here.

1. ~ Orthogonality of ${\bf \nabla} \eta$ and ${\bf \nabla} \chi$ makes it transparent to introduce additional current density from
Eq.(65) that we write as
\begin{equation}
\tilde{\bf J}_s = - \frac{\hbar^2}{2m} {\bf \nabla} (\eta \chi) 
\end{equation}
The velocity field defined by expression (63) is curl-free implying that the vorticity is zero. A new velocity field is defined
using the current density (69): we have two choices $\frac{\tilde{\bf J}_s}{\tilde{\rho}}$ and $\frac{\tilde{\bf J}_s}{\tilde{S}}$.
In either case there exists nonvanishing vorticity indicating rotation. The presence of two scalar fields reminds us the role
of Clebsch potentials in rotating fluid theory.

2.~ The orthogonality condition is not unusual or exceptional; an interesting illustrative example is as follows. Consider
uniform magnetic field along z-direction ${\bf B} = B_0 \hat{z}$. Since ${\bf B} = {\bf \nabla} \times {\bf A}$, assuming
${\bf A} = B_0 \psi {\bf \nabla} \phi$ choosing $\psi = x, \phi = y ; \psi =-y, \phi =x; \psi = x^2/2, \phi = y/x$ we get respectively
${\bf A} = x \hat{y} ; {\bf A} = - y \hat{x} ; {\bf A} = -\frac{y}{2} \hat{x} + \frac{x}{2} \hat{y}$. In all the three cases,
we get the uniform magnetic field and ${\nabla} \psi$ is orthogonal to ${\bf \nabla} \phi$.

Traditionally in the Schroedinger theory there is nothing like spin, however it has been argued in \cite{5} that logically
Schroedinger particle  has to be considered in a spin eigenstate. It is of interest to compare present result with that given in
\cite{5}. The wavefunction (54) is complex, we decompose it into real and imaginary parts and using the rule (6) transform it to real 
wavefunction
\begin{equation}
 \Psi_s = \begin{bmatrix} \psi^0_{sr} + i \psi^0_{si} \\ 0 \end{bmatrix} = \begin{bmatrix} \psi^0_{sr} \\ - \psi^0_{si}\end{bmatrix}
\end{equation}
If complex wavefunction is used we obtain
\begin{equation}
 \Psi^\dagger_s \sigma_3 \Psi_s = {\psi^0_{sr}}^2 + {\psi^0_{si}}^2
\end{equation}
On the other hand, the real wavefunction in (70) leads to the value ${\psi^0_{sr}}^2 - {\psi^0_{si}}^2$. This value differs from (71)
but it is in accordance with the expression (64) suggested here. For the sake of completeness we give the calculated values
for other matrix elements
\begin{equation}
 \begin{bmatrix} \eta & \chi \end{bmatrix} \sigma_1 \begin{bmatrix} \eta \\ \chi \end{bmatrix} = 2 \eta \chi
\end{equation}
\begin{equation}
 \begin{bmatrix} \eta & \chi \end{bmatrix} \sigma_2 \begin{bmatrix} \eta \\ \chi \end{bmatrix} = 2 \tilde{\rho}
\end{equation}
\begin{equation}
 \begin{bmatrix} \eta & \chi \end{bmatrix} C \begin{bmatrix} \eta \\ \chi \end{bmatrix} = 0
\end{equation}
Note that the Pauli matrix $\sigma_2$ in real representation is just the identity matrix $I_2$. 

\section{\bf New perspective on Weyl, Dirac and Majorana equations}

{\bf V-A. ~ Formal Aspects}

Relativistic spinor equations of Weyl and Dirac are first-order space-time derivative wave equations for complex wavefunctions/fields.
The presence of $i$ in Pauli spin matrix $\sigma_2$ and correspondingly in the Dirac matrix $\gamma_2$ is responsible for the complex
nature of the wave equations. Sigma matrices and Dirac gamma matrices satisfy Pauli and Dirac algebras respectively. Is it possible
to have real wave equations for the spinors satisfying consistent (transformed) algebra for real matrices? In the Dirac equation (18)
 $\gamma^0, \gamma^1, \gamma^3$ being real and $\gamma^2$ imaginary, the combination $i \gamma^\mu$ gives a mixed representation.
Replacing all of the gamma matrices by pure imaginary matrices the terms $-i \gamma^\mu_M$ result into a real
wave equation. The new matrices satisfying the Clifford algebra were obtained by Majorana \cite{22} and his equation could be
written as
\begin{equation}
  (\hbar \gamma^\mu_M \partial_\mu ~ - ~ m c ) \Psi_M = 0
\end{equation}

Majorana equation (75) describes a spin-half particle that is its own antiparticle, and electrically neutral. To derive this
equation Majorana considers self charge-conjugate 4-spinors where the charge conjugator matrix $\Gamma$ is
\begin{equation}
 \Gamma = \begin{bmatrix} O & C \\ C^T & O \end{bmatrix}
\end{equation}
Here the null-matrix is denoted by $O = \begin{bmatrix} 0 & 0 \\ 0 & 0 \end{bmatrix}$.
In the convention adopted for the gamma-matrices in \cite{2} the charge conjugate matrix is defined to be $i \gamma^1 \gamma^3 \gamma^4$ that gives (76).
A unitary matrix $S$ is used to change the spin frame $\Psi \rightarrow S \Psi; ~ \gamma^\mu \rightarrow S \gamma^\mu S ^{-1}$ and 
\begin{equation}
 S \Gamma \bar{S}^{-1} = I
\end{equation}
\begin{equation}
  S = \frac{e^{ i \pi / 4}}{\sqrt{2}} (I - i \Gamma)
\end{equation}

We denote $4\times 4$ identity matrix by $I$. One gets pure imaginary representation of 
gamma matrices, and real Dirac wave functions. Corson gives explicit Majorana
matrices \cite{2} though his choice for the unitary matrix $S$ differs from that of Majorana.

There are two novel aspects in our approach: the imaginary unit $i(-i)$ is treated equivalent to the real matrix $C(C^T)$, and the Pauli spin
matrix $\sigma_2$ is equivalent to the identity matrix $I_2$. It immediately follows that the Weyl equation is already a real wave equation
replacing $\sigma_2 \rightarrow I_2$
\begin{equation}
 -c [ \sigma_1 \frac{\partial}{\partial x} + I_2 \frac{\partial}{\partial y} +\sigma_3 \frac{\partial}{\partial z}] \Psi_w\prime
 = I_2 \frac{\partial}{\partial t} \Psi_w\prime
\end{equation}
The unusual matrix algebra due to the commuting matrix $I_2$ makes the interpretation of Eq.(79) intricate; here we have
$\{ \sigma_1, \sigma_3\} =0 ~; \sigma_1^2 =\sigma_3^2 =I_2^2 =I_2$, therefore, the second-order wave equation obtained from (79) using
the operator $[ \sigma_1 \frac{\partial}{\partial x} - I_2 \frac{\partial}{\partial y} +\sigma_3 \frac{\partial}{\partial z}
-\frac{I_2}{c}\frac{\partial}{\partial t}] [ \sigma_1 \frac{\partial}{\partial x} + I_2 \frac{\partial}{\partial y} +\sigma_3 \frac{\partial}{\partial z}
+\frac{I_2}{c}\frac{\partial}{\partial t}] =\frac{\partial^2}{\partial x^2} -\frac{\partial^2}{\partial y^2} + \frac{\partial^2}{\partial z^2}
-\frac{1}{c^2} \frac{\partial^2}{\partial t^2}$ is a non-standard one. It is obvious that Eq.(79) has problem with the Lorentz covariance.

In the case of the Dirac equation (18) one of the gamma-matrices, namely $\gamma^2$ is imaginary and others are real; however $\sigma_2 \rightarrow I_2$
implies that all the 4 gamma-matrices are real. Multiplication with $-i$ gives pure imaginary matrices
\begin{equation}
 \gamma^\mu_d\prime = - i \gamma^\mu
\end{equation}
\begin{equation}
 \gamma^\mu_r = \begin{bmatrix} I_2 & O \\ O & -I_2 \end{bmatrix}, ~ \begin{bmatrix} O & \sigma_1 \\ -\sigma_1  & O \end{bmatrix}, ~
 \begin{bmatrix} O & I_2 \\ -I_2 & O \end{bmatrix}, ~ \begin{bmatrix} O & \sigma_3 \\ -\sigma_3 & O \end{bmatrix}
\end{equation}
Substituting (80) and (81) in the Dirac equation (18) we get the real wave equation
\begin{equation}
 (\hbar \gamma^\mu_d\prime \partial_\mu - m c ) \Psi_d\prime = 0
\end{equation}
The properties of the transformed gamma-matrices differ from the Dirac gamma-matrices that satisfy
 \begin{equation}
  \{ \gamma^\mu, \gamma^\nu \} = 2 g^{\mu \nu}
 \end{equation}
Though we have
\begin{equation}
 -(\gamma^0_d\prime)^2 = (\gamma^i_d\prime)^2 =I
\end{equation}
following relations violate (83)
\begin{equation}
\{ \gamma_r^1, \gamma^2_r \} = -2 \begin{bmatrix} \sigma_1 & O \\ O & \sigma_1 \end{bmatrix}, ~ \{\gamma^1_r, \gamma^3_r \} = -2 
\begin{bmatrix} \sigma_3 & O \\ O & \sigma_3 \end{bmatrix}
\end{equation}
Commutator relations may also be noted $[\gamma^1_r, \gamma^2_r ] = [\gamma^3_r, \gamma^2_r] =0$. Once again we find that the real 
representation of the Dirac equation (82) is at odds with the standard Lorentz covariance prescription.

Formal mathematical manipulations carried out here are correct, therefore, the challenging question is whether the proposed
formal replacements are disallowed on physical grounds or whether there lies deep physics behind them. First we note two remarkable
results obtained using the present approach: derivation of a 2-spinor massless real wave equation in $2+1~D$, and Majorana
equation (75).

In the pseudo-Euclidean space the matrices associated with the basis vectors are real given by the set (32). It is straightforward to
write the 2-spinor wave equation using them
\begin{equation}
[C^T \frac{\partial}{\partial t} + c \sigma_1 \frac{\partial}{\partial x} +  c \sigma_3 \frac{\partial}{\partial y}] \Psi_{w_2} = 0
\end{equation}
It is easy to show that the matrix-algebra (33)-(36) leads to the standard second-order wave equation
\begin{equation}
[-\frac{\partial^2}{\partial t^2} + c^2 \frac{\partial^2}{\partial x^2} + c^2 \frac{\partial^2}{\partial y^2}] \Psi_{w_2} =0
\end{equation}
Note that Eq.(86) differs from the usual $2+1 ~D$ Weyl equation.

To derive Majorana equation we make the replacement $ 0 \rightarrow O, ~ 1 \rightarrow I_2, ~ i(-i) \rightarrow C(C^T)$ in the Pauli
spin matrices to obtain
\begin{equation}
\sigma_1 \rightarrow ~ \begin{bmatrix} O & I_2 \\ I_2 & O \end{bmatrix}
\end{equation}
\begin{equation}
\sigma_2 \rightarrow ~ \begin{bmatrix} O & C^T \\ C & O \end{bmatrix}
\end{equation}
\begin{equation}
\sigma_3 \rightarrow ~ \begin{bmatrix} I_2 & O \\ O & -I_2 \end{bmatrix}
\end{equation}
Multiplying (88) -(90) by $-i$ pure imaginary Majorana matrices in one of the useful representations are obtained
\begin{equation}
\gamma^1_M = - i  \begin{bmatrix} O & I_2 \\ I_2 & O \end{bmatrix}
\end{equation}
\begin{equation}
\gamma^2_M = - i \begin{bmatrix} I_2 & O \\ O & -I_2 \end{bmatrix}
\end{equation}
\begin{equation}
\gamma^3_M = - i  \begin{bmatrix} O & C^T \\ C & O \end{bmatrix}
\end{equation}
Assuming in addition to them the fourth matrix
\begin{equation}
\gamma^4_M = - i \begin{bmatrix} O & -\sigma_1 \\ \sigma_1 & O \end{bmatrix} 
\end{equation}
we get the standard Majorana equation (75). Note that Corson \cite{2} uses the notation $\gamma^1, \gamma^2, \gamma^3,\gamma^4$
for Dirac matrix indices.

{\bf V-B. ~Physical Interpretation}

Real wave equations derived here are new but apparently there are unphysical elements in them, for example, inconsistency with the 
Lorentz invariance. To understand the nature of the problem we examine the role of factorization. In our approach $i^2 =-1$
and $(-i)(i) =1$ with the requirement that we represent the factored components in terms of real numbers leads to the interesting
consequence that a $2 \times 2$ matrix $C (C^T)$ is the simplest representation. This unravels a new feature in the Schroedinger equation
that the presence of $i$ is associated with the spin state of a Schroedinger particle.

The idea of factorization dates back to Schroedinger's derivation of his equation \cite{23}. To explain the main idea let us 
consider the following fourth-order equation for the vibrating plate
\begin{equation}
 [\nabla^2 \nabla^2 + \frac{\partial^2}{\partial t^2} ] u =0
\end{equation}
Schroedinger found fourth-order wave equation for the scalar field and drew attention to Eq.(95) in a footnote in \cite{23}. Since
$\nabla^2 = {\bf \nabla}.{\bf \nabla}$ is a scalar operator one can factorize the bracketed operator in (95) using the imaginary unit
$i$:  $(\nabla^2 + i \frac{\partial}{\partial t})(\nabla^2 - i \frac{\partial}{\partial t})$. Schroedinger employed this method to
derive his famous equation.

The second-order massless wave equation (1) cannot be factorized using this method since ${\bf \nabla}$ is a vector operator. Introducing
sigma-matrices the scalar product ${\bf \nabla}.{\bf \nabla}$ can be re-written as an ordinary product of scalars
\begin{equation}
\nabla^2 ~ \rightarrow ~ ({\bf \sigma}.{\bf \nabla})({\bf \sigma}.{\bf \nabla}) ~ \rightarrow ~ \nabla^2 I_2
\end{equation}
The factorization now demands a two-component spinor instead of a scalar wavefunction, and spin appears in the manifest form in the 
Weyl equation (15). Imaginary unit $i$ is not explicitly present in (15) since the factorization is just
$(c {\bf \sigma}.{\bf \nabla} -\frac{\partial}{\partial t} I_2)(c {\bf \sigma}.{\bf \nabla} +\frac{\partial}{\partial t} I_2)$. The 
most celebrated, of course, is Dirac's factorization for the KG equation (3). Now one needs $4 \times 4$ matrices to affect factorization of
the operator in the form $(i \hbar \frac{\partial}{\partial t} + i \hbar c {\bf \alpha}.{\bf \nabla} - \beta m c^2)
(i \hbar \frac{\partial}{\partial t} - i \hbar c {\bf \alpha}.{\bf \nabla} + \beta m c^2)$. The KG wavefunction must be replaced by a
4-component Dirac spinor in the Dirac equation. Note that both Dirac and KG equations are relativistic quantum wave equations; the new
feature of spin in Dirac equation is a consequence of factorization/linearization.

The factorization seems to reveal new physics, and also brings out puzzling questions as exemplified by Schroedinger and Dirac equations.
Apparent unphysical consequences have to be re-analyzed in the light of this. The main problem arises
from the factorization of $\sigma_2$ given by Eq.(37). To address this problem, the role of $i$, the significance of the phase factor in the 
state eigenvectors, and the meaning of the commutation relations of the Pauli matrices are explored in the topological perspective. Since
state vectors and Heisenberg's canonical commutation relations are two of the cornerstones of quantum theory, we approach this question via these
two routes, and arrive at the important result that spin has a topological origin.

{\bf Two-state system and phase}

In the standard theory of Pauli matrices only one of the sigma-matrices, 
conventionally chosen to be $\sigma_3$, could be diagonal. The orthonormal eigenvectors for the eigenvalues $\pm 1$ are given by
\begin{equation}
\sigma_3 \begin{bmatrix} 1 \\ 0 \end{bmatrix} = \begin{bmatrix} 1 \\ 0 \end{bmatrix} ; ~
\sigma_3 \begin{bmatrix} 0 \\ 1 \end{bmatrix} = -1 \begin{bmatrix} 0 \\ 1 \end{bmatrix}
\end{equation}
Matrices $\sigma_1$ and $\sigma_2$ also have eigenvalues $\pm 1$. Their orthonormal eigenvectors are
\begin{equation}
 \sigma_1 \frac{1}{\sqrt 2} \begin{bmatrix} 1 \\ 1 \end{bmatrix} =\frac{1}{\sqrt 2}\begin{bmatrix} 1 \\ 1 \end{bmatrix};
 ~\sigma_1 \frac{1}{\sqrt 2} \begin{bmatrix} 1 \\ -1 \end{bmatrix} =-\frac{1}{\sqrt 2} \begin{bmatrix} 1 \\ -1 \end{bmatrix}
\end{equation}
\begin{equation}
 \sigma_2 \frac{1}{\sqrt 2} \begin{bmatrix} 1 \\ i \end{bmatrix} =\frac{1}{\sqrt 2}\begin{bmatrix} 1 \\ i \end{bmatrix};
 ~\sigma_2 \frac{1}{\sqrt 2} \begin{bmatrix} i \\ 1 \end{bmatrix} =-\frac{1}{\sqrt 2} \begin{bmatrix} i \\ 1 \end{bmatrix}
\end{equation}

To delineate the role of $i$ and phase factors we first consider the action of $\sigma_1,~\sigma_2,~C^T$ on the spin-up and spin-down states (97)
\begin{equation}
 \sigma_1 \begin{bmatrix} 1 \\ 0 \end{bmatrix} = \begin{bmatrix} 0 \\ 1 \end{bmatrix}; ~
 \sigma_1 \begin{bmatrix} 0 \\ 1 \end{bmatrix} = \begin{bmatrix} 1 \\ 0 \end{bmatrix}
\end{equation}
\begin{equation}
 \sigma_2 \begin{bmatrix} 1 \\ 0 \end{bmatrix} = e^{i\pi/2}\begin{bmatrix} 0 \\ 1 \end{bmatrix}; ~
 \sigma_2 \begin{bmatrix} 0 \\ 1 \end{bmatrix} = e^{-i\pi/2} \begin{bmatrix} 1 \\ 0 \end{bmatrix}
\end{equation}
\begin{equation}
 C^T \begin{bmatrix} 1 \\ 0 \end{bmatrix} = \begin{bmatrix} 0 \\ 1 \end{bmatrix}; ~
 C^T \begin{bmatrix} 0 \\ 1 \end{bmatrix} = e^{i\pi} \begin{bmatrix} 1 \\ 0 \end{bmatrix}
\end{equation}
Eqs. (100)-(102) show that all the operators flip the spin states with the only difference being in the accompanied phases. In fact, rotations
about appropriate axes and through suitable angles map various eigenvectors onto each other. For example, a rotation through an angle
$-\pi/2$ about x-axis maps the eigenvectors (97) of $\sigma_3$ onto the eigenvectors (99) of $\sigma_2$. An interesting case is that of
rotation about y-axis using the unitary matrix
\begin{equation}
 U = \frac{1}{\sqrt{2}} (I_2 + C^T)
\end{equation}
The matrix (103) transforms spin-up along z-axis to spin-up along x-axis.

In quantum theory usually phase transformed wavefunction does not affect the physical observables. The nontrivial geometric phases a la Aharonov-Bohm
and Berry phases, have been demonstrated experimentally. Geometry of quantum state space provides nice explanation of the geometric phases. However
here we are interested in real wave equations. Hestenes has developed a mathematical language that he terms geometric algebra (GA). Imaginary
unit $i$ has a geometric interpretation and spinors are real in GA version of ``Real Quantum Mechanics''. Unfortunately, in spite of some
novel aspects GA has remained a sterile alternative to the standard quantum mechanics \cite{24}. Note that the argument that Schroedinger particle
is in a spin eigenstate \cite{5} does not depend on GA formalism. Proposed interpretation of $i$ beyond geometry in terms of a topological
obstruction in 1D directed real line presented in section III-B marks a radical departure. Let us return to the $Z_2$ vortex problem.

Motivation for the abstract Hamiltonian (45) is historically related with the problem of poly-atomic molecules studied by Herzberg and 
Longuet-Higgins \cite{25}. Authors point out that the topology of conical intersection is responsible for the sign-change in an electronic
wavefunction governed by the Hamiltonian (45). Wilczek brings out the significance of geometric matrix or Berry phase in this connection \cite{4}.
For $Z_2$ vortex the gauge potential is calculated given by (50). Path-ordered integral defining the geometric matrix in \cite{4} does not 
contain $i$ that appears now in the expression (50). Multiplication by scale factor for cylindrical geometry finally gives the value of the pure gauge
potential to be $-\frac{1}{2 r}$ in geometrical unit. This azimuthal vector potential should not be confused with the magnetic vector potential
in the field-free region of actual (confined) magnetic flux in the Aharonov-Bohm case.

Recently modular angular momentum exchange as a physical mechanism for the origin of Aharonov-Bohm effect has been proposed \cite{26}
in which the pure gauge potential is suggested to have angular momentum $\frac{e}{c} {\bf r} \times {\bf A}$. A logical extension of this idea 
to the $Z_2$ vortex then associates dimensionless angular momentum of magnitude $-\frac{1}{2}$ with the azimuthal vector potential here.
Angular momentum unit is arbitrary, assuming it to be $\hbar$ we get spin-half. We have arrived at a remarkable result: the sign-change
in the real spinor (48) upon $2\pi$ rotation and topological origin of intrinsic spin $\frac{\hbar}{2}$ are inter-related. Unitary
transformations to implement phase changes among various eigenvectors (97) to (102) seem to signify the role of geometric vector
potential (50). Note that the only complex sigma-matrix could be put in a suggestive form
\begin{equation}
 S_2 = \frac{\sigma_2}{2} \hbar = -A_\phi \hbar C^T
\end{equation}

{\bf Pauli algebra and topology}

It seems complex wave equations tend to obscure the physical origin of spin \cite{5, 17}. It is of interest to investigate this issue in the context
of the Lie algebra of SU(2) group; it is the Pauli algebra
\begin{equation}
 [ \sigma_i, \sigma_j ] = 2 i \epsilon_{ijk} \sigma_k
\end{equation}
For the real matrix operators we have obtained the algebra given by (34) to (36) making use of the transformation (6) and pseudo-Euclidean
space. Surprisingly the algebra (105) is identical to real commutators (34) to (36) since the imaginary unit $i$ cancels out in (105).
The obvious question is about the meaning of the commutators. Interpretation of $i$ suggested here, see section III-B, identifies point defect 
on real 1D directed line to the nontrivial topology. It is natural to ask whether this proposition could be extended to the directed
2D area. Here we expect a surface discontinuity when a unit normal to the surface reverses direction. If product of two sigma-matrices
represents directed area element then the right hand side of Eq.(105) represents the topological deficit/defect. Recall that the 
Pauli algebra (105) is just spin operator commutator in quantum mechanics
\begin{equation}
 [S_i, S_j ] = i \hbar \epsilon_{ijk} S_k
\end{equation}
the area element corresponds to phase space of canonical variables momentum and position.

Arbitrarily chosen direction along z-axis, and two projections of spin $\pm \frac{1}{2} \hbar$ show that the proposed topological
obstruction lies in x-y plane, and the corresponding area element is that of phase space. The role of imaginary unit $i$ in the algebra
(105) and the Weyl wave equation (15) is that of continuous rotation around the given axis. On the other hand, real wave equation (79) incorporates
nontrivial topology in the replacement $\sigma_2 \rightarrow I_2$. Intuitively the difference is akin to the imaginary time coordinate
in the Minkowskian geometry and real time coordinate in the pseudo-Euclidean geometry of space-time: Lorentz invariance is not violated
in either case the only difference is the contravariant/covariant description in the later. For a manifest Lorentz covariance in the
new real wave equations (79) and (82) further work is needed.

\section{\bf Outlook on particle physics}

Symmetry-inspired classification of the observed elementary particles is founded on space-time symmetries as well as internal symmetries
associated with the hypothetical spaces. Unified theory of fundamental interactions, excluding gravity, is also based on the postulated
internal space gauge symmetries $SU(3)_c \times SU(2)_L \times U(1)_Y$ in the standard model (SM) of particle physics. The question on
``Beyond SM'' has also acquired a sort of standard form \cite{12}. However there is a growing feeling among particle physicists that
some crucial thing is missing to arrive at a reasonably complete picture of physical reality. Could such an idea come discarding space-time
reality? Influential school of thought in the mainstream explores this line of thinking. We believe that space-time has intuitive
perception; moreover energy, momentum, and angular momentum conservation laws having unequivocal experimental validity, at least, indirectly
provide support to the physical reality of space-time. The biggest stumbling-block in the approach assuming geometry and topology of
space-time being fundamental, is that electric charge, weak charge and color charge are not related with any kind of space-time symmetry.
Here it is shown that it is possible to explain the meaning of coupling charges with space-time symmetry.

{\bf VI-A ~ Electron Model: General Considerations}

Founders of QED, for example, Dirac and Pauli were critical of the theory mainly regarding the mathematical handling of infinities in the 
renormalization method. Note that QED is a paradigm for SM. In Pauli's opinion the determination of the fine structure constant,
$\alpha = e^2/\hbar c$, was a fundamental problem. Feynman also advocated ``more insight and physical intuition'' in QED calculations. We refer
to a recent discussion in \cite{27}. We have suggested that the simplest and the earliest discovered elementary particle, namely the electron,
holds the key to unravel the mysteries of the particle world \cite{28}. Unlike Dirac, our question is what makes the renormaliztion
procedure work so successfully. Unlike Pauli, rather than the number 137, we are intrigued by the composition of $\alpha$. One could
view $\alpha$ as a length ratio of classical electron charge radius $a_e = \frac{e^2}{m c^2}$ and Compton wavelength $\lambda_c =\frac{\hbar}{m c}$;
as a ratio of two angular momenta $e^2/c$ and $\hbar$; and as a ratio of flux quanta $2\pi e$ and $h c /e$. Do such speculations have
physical significance? In the past we have investigated this question and found some new physical ideas; recently the electron magnetic moment
calculated in QED, $\mu_e$ has been re-interpreted geometrically \cite{27}. The main idea follows from the re-written form of the expression
of $\mu_e$ calculated to the second order in $\alpha$
\begin{equation}
 \mu_e = \frac{e}{m c} [ \frac{\hbar}{2} + \frac{f}{2} - 0.328478444 \frac{\alpha^2}{\pi^2}\frac{\hbar}{2}]
\end{equation}
where $f=e^2/2\pi c$. Feynman's challenge on a physically intuitive method of computation of individual terms in (107) including the sign
has drawn attention of physicists \cite{29}. Our concern is markedly different than Feynman's challenge. We propose that the decomposition
(107) throws light on the electron structure in view of the following points. 1) The bracketed terms comprise of only fundamental constants
$e, \hbar, c$. It is important to note that muon magnetic moment also has identical first two terms. The third term differs in sign
and numerical value due to vacuum polarization and muon being much heavier than electron. Since muon is unstable decaying to electron
we attach fundamental significance to only the electron magnetic moment expression. 2) It is fascinating to observe that $e^2/c$ has the
dimension of angular momentum \cite{28}. Expressing the electromagnetic quantities in the geometrical units it is this ratio that occurs
in the Lorentz force and the generalized momentum while $e$ cancels out in the Maxwell field equations \cite{27, 28}. And, 3) The 
magnetomechanical ratio, in the units of $\frac{e}{2mc}$ has a profound classical explanation for the orbital motion of an electron.
Spin hypothesis assumes the magnetomechanical relationship but this ratio, i. e. g-factor is twice to that of the orbital motion. A
natural extension of the spin hypothesis would mean interpreting the bracketed term in (107) as a generalized spin series
\begin{equation}
 \bar{S_e} = \frac{\hbar}{2} + \frac{f}{2} - 0.328478444 \frac{\alpha^2}{\pi^2}\frac{\hbar}{2}
\end{equation}

This conceptual framework led to the development of a multi-vortex internal structure of the electron in a nontrivial geometry of
cylindrical space-time \cite{27}. Advancement in the electron model is envisaged in the light of the present work. In the standard 
theory the first term in (108) represents SAM of a double-valued spinor (97) for a chosen fixed z-direction with projections $\pm \hbar/2$.
Intrinsic spin, non-describable classically, is an internal dynamical variable for the assumed point electron. First let us note that
for a massless spin-half particle one has the Weyl equation (15) satisfied by a 2-spinor and $\hbar$ does not appear in this equation. 
Following Eq.(16) we suggest that the magnitude of SAM is arbitrary, and could be determined by the variable $y$. Logically we may
associate double-valued spinors with each term in the series (108). Thus the first notable result is that expression (108)
represents three 2-spinors with different SAM. If there is any internal motion or space-time structure for spin, it has to be in the transverse
x-y plane. Multi-vortex model satisfies this condition \cite{27}, and with the new result each vortex has a corresponding 2-spinor.

Obviously physical electron is not massless and also has charge $-e$. Simplified cylindrical space-time geometry and isolated multi-vortex
multiple spinors picture has to be changed. An attractive idea is to explore the role of a logarithmic spiral defined by $r=r_0 e^{p\theta}$,
where $r_0$ is a constant and $p$ is a real number. Note that a circle is a spiral with $p=0$, therefore, multiple circles connected in a subtle
way could be a realistic geometric picture of the electron. In $r-\theta$ plane the rate of growth of the spiral is determined by $p$, and 
a cycle or a winding or a turn is obtained varying $\theta$ from $0 \rightarrow 2\pi$. Logarithmic spiral has many interesting symmetry
properties; we mention a few : $p \rightarrow -p$ gives a mirror image of the spiral; for $r_0 =1$, $r \rightarrow 1/r, ~ \theta \rightarrow \theta$
maps the spiral to its mirror image; and, every straight line through pole intersects the spiral at the same angle. To construct electron model
we incorporate additional feature of time dependence and modification due to a postulated topological obstruction in the spiral. Time-dependent
spiral is obtained by assuming $\theta =\omega t$ in analogy to a circular motion. The continuous spiral curve is modified to a collection
of approximately circular patches when a jump occurs at the crossing $\theta =2\pi$; see section III-B.
The patches are connected in a subtle manner. Now we have
three parameters $r_0, p, \nu=1/T$ that define the logarithmic spiral; the jump discontinuity may be attributed either to $\nu$ or to
time itself.

A 2-spinor corresponding to each term in the expression (108) is defined on each patch, and the connection is suggested to be of Aharonov-Bohm like pure
gauge potential of the kind discussed for $Z_2$ vortex in section III-C. To be specific let us consider first two terms in (108). A qualitative
picture may be built assuming $r_0 = \lambda_c$, and $p<<1$. At $\theta = 2\pi$ expanding the exponential in power series up to the second term
$r\approx r_0 +2\pi p r_0$. Taking $p= e^2/h c$, the second term becomes equal to the classical electron charge radius $a_e$. Frequency is another
parameter, assuming it to be equal to $m c^3/e^2$, 137 turns of the spiral would occur in the time period $T_c =\hbar/m c^2$ corresponding to 
the Compton wavelength. The spiral in the $r-\theta$ plane propagates along z-direction; suppose this motion is also periodic with time
periodicity $T_c$. Then there are 137 turns of the spiral for each time period, and in the series (108) one may expect 137 sequences of patches with
decreasing SAM. The mirror image of the electron spiral structure may be identified with positron, and $t \rightarrow -t$ transformation 
in the internal motion transforms the spiral to its mirror image.

The speculated geometric picture of the electron has an element of reality since empirical parameters $\lambda_c, a_e$ have been used. 
Mass appears in both lengths. Decoupled Dirac equation (18) for $m=0$ is transparent in Weyl or chiral representation (19)
in which  Dirac gamma matrices are $\gamma^0 =  \begin{bmatrix} O & I_2 \\ I_2 & O \end{bmatrix};\gamma^i = 
\begin{bmatrix} O & \sigma_i \\ -\sigma_i & O \end{bmatrix}$. Since $\hbar$ is not present in the massless 2-spinor equation, we
are free to put by hand $\hbar, f$ in the following equations
\begin{equation}
 i\hbar(\frac{\partial}{\partial t} + c {\bf \sigma}.{\bf \nabla}) \Psi_R =0
\end{equation}
\begin{equation}
  i f(\frac{\partial}{\partial t} - c {\bf \sigma}.{\bf \nabla}) \Psi_L =0
\end{equation}
It is proposed that $\Psi_R, \Psi_L$ are coupled in the following way
\begin{equation}
 i\hbar(\frac{\partial}{\partial t} + c {\bf \sigma}.{\bf \nabla}) \Psi_R =m c^2 cos \alpha_0 \Psi_L
\end{equation}
\begin{equation}
i f(\frac{\partial}{\partial t} - c {\bf \sigma}.{\bf \nabla}) \Psi_L =m c^2 sin \alpha_0 \Psi_R
\end{equation}
The coupling angle is assumed small equal to $p=e^2/hc$, then we may take $cos \alpha_0 \approx 1; ~sin\alpha_0 \approx \alpha_0$. In this case
combined equations (111) and (112) represent just the Dirac equation (18).

Mechanical interpretation of electronic charge in terms of spin $f$ suggested earlier \cite{27, 28} could be given more secure
foundation using the factorization and real representation. The first two terms in (108) could be factorized in either of the two forms 
given below
\begin{equation}
 \frac{\hbar}{2} + \frac{f}{2}= \frac{1}{4\pi c}(\sqrt{hc} +ie)(\sqrt{hc} -ie)
\end{equation}
\begin{equation}
 \frac{\hbar}{2} + \frac{f}{2}= \frac{1}{4\pi c}(e+i\sqrt{hc}) (e-i\sqrt{hc})
\end{equation}
In the light of the smaller length scale corresponding to $f/2$ the factorization (113) is more plausible than the one given by (114).
The phasor form $\sqrt{hc} +ie \rightarrow ~ \sqrt{hc +e^2} exp(i \arctan \frac{e}{\sqrt{hc}}) \approx \sqrt{hc+e^2} exp(i\frac{e}{\sqrt{hc}})$
indicates why U(1) plays important role in QED. Interpreting $hc$ analogously to $e^2$, a charge-like interpretation follows for
$\sqrt{hc}$. Let us denote this charge by $e_s$ then a fine structure or coupling constant may also be defined for this charge
\begin{equation}
 \alpha_s^e = \frac{e_s^2}{\hbar c} = 2 \pi
\end{equation}
Now recalling that the running coupling constant in QCD has the value of the order of $\alpha_s^e/2\pi$ \cite{12} it is tempting to
identify it with the strong interaction. It would certainly be a daring or crazy idea to relate spin $\hbar/2$ with strong interaction,
nevertheless rather than rejecting it outright we proceed further to examine its consequences.

One of the main reasons for magnetic monopole hypothesis is the asymmetry in the sources in the Maxwell equations \cite{28}. Quantization
of charge-monopole system shows that the monopole fine structure constant is $\alpha_m = 137 ~(137/4)$. Extensive experimental searches
carried out over decades have failed to detect a monopole. The huge coupling constant $\alpha_m$ also shows no signature in the nature.
In contrast, it is for historical and chronological reasons that Maxwell equations could not incorporate the spin of the electron. Rather
than a point electron charge, the physical attributes of 2-spinors for the first two terms in the expression (108) corresponding to
charges $e_s, e$ respectively would bring the desired symmetry in the suitably modified Maxwell equations. The monopole hypothesis would
become superfluous in that case. Not only this, the coupling constant (115) falls in the range of physically observed interactions. 
In fact, following the present logic, the third term in (108) needs to have charge-like interpretation; denoting it by $e_w$ the coupling
constant is of the order of $10^{-5}$. The value of the dimensionless Fermi coupling constant $G_F M_p^2$ setting energy scale
at the proton mass $M_p$ is also of the same order. Is this coincidence accidental? It may be suggested that this explains weak
interaction consistent with the present approach.

Spin-charge connection may be approached from another angle using C-matrix, we have
\begin{equation}
 \sqrt{hc} + i e \rightarrow ~ \begin{bmatrix} \sqrt{hc}  & e \\ -e & \sqrt{hc} \end{bmatrix}
\end{equation}
The eigenvalues of the matrix (116) are complex $(\sqrt{hc} \pm i e)$. A curious observation in section 20 of \cite{2} is worth mentioning.
Gauge invariance of complex scalar field Lagrangian density leads to a conserved Noether current, and interaction with electromagnetic
field may be understood treating this current as electromagnetic current. Alternatively, the gauge transformation may be viewed as
a rotation in 2D ``symbolic isotopic-spin space'' such that the conserved quantity is now angular momentum in this space.

To summarize, we have shown that spin has topological origin in space-time, and charge is connected with spin; it may be stated in the form
of spin origin of charge (SOC) hypothesis. SOC has two remarkable consequences: a unified picture of strong, electromagnetic and weak
interactions already exists embodied in the expression (108), and spin and charge being related with space-time geometry and topology
the postulated internal spaces in SM become superfluous or artefacts.

{\bf VI-B ~ QCD and Unification}

The most serious objection that immediately arises is that electron is believed to have no strong interaction that sharply contradicts
SOC. The resolution on this question is sought along two lines. First we re-visit high energy scattering experiments. Landmark
MIT-SLAC $e^- p$ deep inelastic scattering (DIS) experiment paved the path for QCD. Another important experiment is that of 
high energy electron-positron annihilation producing hadrons in the final state $e^-e^+\rightarrow hadrons$ ;
section 5.2 of \cite{12} mentions this experiment as evidence for QCD. This experiment was, in fact, a part of November
Revolution famous for the discovery of $J/\Psi$ resonance interpreted as $c\bar{c}$ bound state. This experiment seems ideal to test SOC since
no hadrons are present in the initial state. We know that QED elementary process is $e^-e^+ \rightarrow 2 \gamma$. At high energy the 
final states are lepton pairs, theory and experiment agree well as shown by the ratio of the cross sections $\sigma(e^-e^+ \rightarrow \tau^-\tau^+)/
\sigma(e^-e^+ \rightarrow \mu^-\mu^+)$ claculated in QED and measured experimentally. Hadrons in the electron-positron annihilation
are viewed as $e^-e^+ \rightarrow q\bar{q} \rightarrow hadrons$. Amazingly just replacing muon charge with quark charge, and counting
the quarks one gets the high energy limit of the ratio of total cross sections, to lowest order in QED, $R= \sigma(e^-e^+\rightarrow hadrons)/
\sigma(e^-e^+ \rightarrow \mu^-\mu^+)$ that agrees with the experiments. Resonances $\phi(s\bar{s}); J/\Psi; \Upsilon(b\bar{b})$ were 
inferred from these maesurements, see Fig.5.6 in \cite{12}.

The counting of color degree of freedom for quarks in the QED calculation is adequate for agreement with measured cross sections, and it
is taken as evidence for QCD. However it is beyond doubt that hadronization is not explained in the theory though the final observed states
are hadrons. Moreover perturbative QCD (pQCD) also has limitations as discussed in \cite{30}. A thorough treatment on the fundamental questions
related with high energy electron-positron annihilation process is given in Chapters 4 and 12. The running coupling constant in QCD 
tends to zero at zero distance, termed as asymptotic freedom, and due to this pQCD and renormalization of ultraviolet divergences could
be justified. Infrared safety comes to the rescue in cancelling divergences in the individual terms giving sensible total cross section
in pQCD for $e^-e^+$ high energy annihilation. Hadronization is a nonperturbative process; Collins suggests the role of experiment,
semiclassical intuition, and lattice gauge theory calculations for this purpose, see section 4.3.1 in \cite{30}. Additional
postulate other than QCD is imperative, for example, 'the breakable string picture' or 'unbreakable elastic spring picture' 
to understand hadronization. Admittedly QCD is the best theory of strong interaction at present, however several important 
questions remain unanswered.  The monograph \cite{30} is not dogmatic \cite{31} and throws insighful light on unsatisfactory issues 
in QCD: understanding hadronization, theory of bound states of quarks, and color confinement hypothesis. It is reasonable to conclude
that the modest success of pQCD in understanding $e^-e^+ \rightarrow q\bar{q}$ allows the exploration of an alternative approach.

In our view, there are two fundamental questions that seem to have been overlooked in the QCD literature. Running coupling
constant implies only changes in the numerical strength with energy; it does not mean changing the nature of charges color
$\leftrightarrow $ electric charge. The second question is that if leptons do not have strong interaction then how does initial
electron-positron state transform to strongly interacting quark pairs. The hypothesis that quarks and gluons do not exist free in
nature saves QCD from embarassment. The claimed discovery of six quarks \cite{32} has to be seen as indirect.

The second line of reasoning is to recognize that SOC radically alters spin-charge relation: half-integral spin and charge(s)
are twin facets of the same physical reality. The magnitude of SAM and numerical strength of the coupling constant vary in strong,
electromagnetic and weak interactions, but qualitatively all the three are similar - a plausible picture is in terms of
vortex-vortex interaction. Thus geometry and topology of space-time vortices would assume fundamental role. Atiyah draws attention to
the significance of the knotted vortex tubes in the atomic model of Lord Kelvin in the 19th century \cite{33}. Note that in physics
literature, specially on Berry phase, quite often geometry and topology are incorrectly used synonymously. Atiyah in a lucid manner
explains their distinction: geometry is quantitative and mostly local in usual physical situations, wheras toplogy is qualitative,
global and discrete. In the proposed scenario electron comprises of mainly three vortices: a core vortex $V_c$ with SAM of $\hbar/2$ 
and charge $e_s$, an orbiting vortex $V_e$ with SAM of $f/2$ and charge magnitude $e$, and a tail vortex $V_w$ corresponding
to the last term in (108). Though the presence of spin is shown in most cases, the charge $e_s$ is hidden or better non-manifest; it 
becomes effective in high energy phenomena. The high energy electron-positron annihilation process offers an opportunity
for re-examination of QCD calculations in the light of this.

The electron model admits a natural subset in which only two vortices $V_c,V_w$ are present; such an object may be identified as
neutrino. Geometrically the electron mass parameter arises due to the orientation of vortices $V_e,V_w$ with respect to the core vortex.
Neutrino is expected to have smaller mass than electron mass since only the weak vortex $V_w$ is involved. The only
fundamental particles are then electron, positron and neutrino which are the constituents of all elementary particles. A qualitative
picture could be built treating them as knots due to their stability and variety \cite{33}. There are no quarks in this picture, however
neutrinos in the knotted structures with electrons and positrons may acquire effective fractional quark-like charges. Just as 
mass of electron is determined by geometry, the mass of elementary particles would be determined by the knot geometry. In the past,
trefoil knot as geometrical model of electron and quarks had been attempted. Trefoil knot is a torus knot $T_{n,m}$ with $(n=2, m=3)$, here
$n$ is the number of windings of the meridian and $m$ is that of the longitude of the torus. In the early days of particle physics, many
physicists were fascinated by the empirical observation that $\alpha^{-1}$ times electron mass multiplied by integers or rational numbers
approximates the masses of some particles, for example, muon mass $m_\mu \approx \frac{3}{2} \alpha^{-1} m$, pion mass 
$m_\pi \approx 2 \alpha^{-1} m$, and Sternheimer \cite{34} in 1968 found at least 14 hadrons having $M\approx integer \times m_\pi$.
Unfortunately these speculations did not have the desired success.

In the light of the new insight on the physical interpretation of spin and charge, SOC hypothesis, and logarithmic spiral geometric
element with the number $p=\alpha/2 \pi$, reviving the past ideas may prove fruitful. For example, Dirac's pulsating spherical shell
model \cite{8} and recent concentric spherical shell model \cite{27} do offer a limited geometric perspective, however the logarithmic
spiral may turn out to be a realistic choice to represent the internal structure of physical particles. In an Archimedian or linear 
spiral each successive turn increases the distance from the pole by a constant difference, whereas the physical length
scales indicate constant ratio making logarithmic spiral more appropriate. We mention that the ratio of atomic Bohr radius $\hbar/m e^2$ to
$\lambda_c$, and  $\lambda_c$ to the electron charge radius is $\alpha^{-1}$. Further, in the mass spectrum of some elementary particles
also appears $\alpha^{-1}$. A qualitative picture for the classification of elementary particles, in analogy to flavor symmetry groups may
emerge utilizing the classification of knots \cite{33}. In particular, it is interesting to note that 3-torus roughly corresponds to a product
of three circles and SU(3); torus knot itself has a great variety.

Nonperturbative QCD and color confinement are the most challenging problems in QCD, and none of the numerous models/mechanisms suggested for them
is satisfactory; see section 5.7 in \cite{12}. Does our model have a potential to be developed as an alternative to QCD? The answer would 
depend ultimately on the quantitative results and viable calculational techniques. At present, we may offer only an outline towards this 
objective keeping in mind that the insightful features in QCD need to be incorporated in the proposed framework. We discuss below two of the
remarkable inputs from QCD.

In the breakable string picture, the collapsed gluon field is visualized as a string or a flux tube with a fixed area of cross section
and uniform energy per unit length; semiclassical Lund string model has been used to explain the hadronization \cite{30}.
Color confinement mechanism in the string picture has found support in lattice QCD \cite{12}. A short comment on infrared slavery in QCD \cite{35}
draws analogy to magnetic monopole confinement in a superconductor in which monopole and antimonopole form a flux tube; in QCD a color
flux tube between quark and antiquark is envisaged. In our model, there does not appear an analogue to gluon, though neutrino 
embedded in a nontrivial geometry may be likened to a quark. Color flux tube is replaced by vortex tube, and a kind of universal
interaction is that between vortices.

Physics of fluid flow and vortex becomes important in our model, and in view of treating a line vortex or even a vortex tube as a string
or strand geometrically, the geometry of knots also becomes important. Atiyah has remarked that in the mid-19th century fluid theory 
was very well developed, and later Tait carried out an extensive tabulation of knots \cite{33}. Today both fluid theory and knot theory
have advanced tremendously, therefore the mathematical formalism of our model is possible but very  difficult: analytical 
treatment of vortex dynamics for more than two is not easy, on the other hand recognition of the new elements proposed in our model
should enable one to simplify the problem. In fact, vortex description has in recent years found many applications including
optics and quantum theory. Optical vortex and quantum vortex are usually viewed as phase vortices, However they are different at a basic 
level since in optics the complex representation is for convenience and dispensable while quantum vortex occurs only for complex
wavefunction and Schroedinger equation itself is complex \cite{36}. Madelung's  hydrodynamical interpretation of quantum theory depends on the 
separation of Schroedinger wave equation into real and imaginary parts. The derivation of real wave equations in this paper makes
it possible to have a natural fluid interpretation and seek vortex solutions. For the purpose of electron model vortex solution in
massless 2-spinor wave equation would be needed. The next question relates to vortex-vortex interaction, and it is not easy as shown
in a monograph \cite{37}. In general, rotation rates, vorticity, energy, angular momentum, number of vortices and the dimensionality of space
determine the emerging patterns and their stability. Spiral structures and coalesce of vortices into a single vortex core, and windings
between cores have both experimental and theoretical support. In our model we can fix certain things: two vortex tubes for neutrino and three
vortices for electron, prescribed SAM given by Eq.(108), and 2D space plus 1D time.

In the geometric picture we need the language of knots that has been used in the past for the vortex atom model \cite{33}. The first
important object is three vortex model of electron: is it a knot, e. g. a torus knot? In an interesting approach called quantum cohomology
Post offers a new perspective on particle physics \cite{38}. Anomalous magnetic moment of electron is explained by him in a trefoil knot model,
rejecting the earlier ring models proposed in the literature. Unfortunately, Post did not ask, and explain the fundamental
question that of the nature of charge, therefore, this model does not go much far. In our electron model, we have to face two issues: 
electron has no visible strong interaction that we attribute to the core vortex $V_c$, and has widespread presence of the electric charge
that we associate with the vortex $V_e$. Is it possible to address these issues in a geometric framework?  The answer in affirmative is
provided visualizing core vortex to form almost a circle, and the vortex $V_e$ as well as $V_w$ to have a $(2+1)$D braid structure.
Though mathematically a circle is a closed curve, in the discussion on topology and $Z_2$ vortex in section III we have introduced a 
discontinuity or defect connecting two patches on the circle; it is for this reason that we envisage an almost circle for the core 
vortex thus admitting strong interaction at short distances. A knot is a closed curve of string with crossings and entangled structures;
a circle is called unknot though it is also a closed curve. A braid unlike the knot has loose ends. Electron and neutrino are proposed
to have almost circle plus braid structure of vortices. The geometrical objects identified with electron, positron and neutrino are the 
building blocks for the various knotted structures, and represent the elementary particles. As a curiousity it is interesting that
Atiyah mentions \cite{33} an interesting $(2+1)$D braid model due to Witten to interpret knot invariant Jones polynomial.

QED as a U(1) gauge field theory is a paradigm for the modern gauge theories. QCD is defined by $SU(3)_c$ gauge invariant Lagrangian
density for quark Dirac 4-spinor fields carrying color and flavor indices, and the nonabelian gluon gauge field. MIT-SLAC DIS experiment
showed point-like constituents of proton that behave as if free at short distances. Almost vanishing coupling constant in the ultraviolet
limit known as asymptotic freedom played a crucial role in the development of QCD \cite{12, 30}. The concept of running coupling constant
depending on the renormalization scale is formulated in the form of renormalization group equation (RGE). For DIS experiment, one may
set the renormalization scale to $Q^2$, where $-Q^2 ={(q - q^\prime)}^2$ with $q, q^\prime$ four-momenta of incident and scattered electron
respectively. For any such kind of high energy process in QCD RGE \cite{12} is
\begin{equation}
 \frac{dg^2_s}{d~ ln Q^2} = 4 \pi \beta(g_s^2) = b g^4_s
\end{equation}
\begin{equation}
 b= -\frac{1}{{(4\pi)}^2}[11-\frac{2}{3} n_f]
\end{equation}
Here the beta function is calculated at 1-loop level, $n_f$ is the number of quark flavors, the gauge coupling constant is $g_s$ and the 
strong fine structure constant in natural units is $\alpha_s = g^2_s/4\pi$. Collins \cite{30} lists $\beta$ function upto 3-loop level
calculations. A lucid discussion on the physical meaning of the beta function, and renormalization group is given in \cite{39}. Note
that negative $\beta$ function in SU(3) gauge group is essential for the asymptotic freedom in QCD.

Running coupling constant in QFT, asymptotic freedom in QCD, and QED fine structure constant $\alpha$ showing increasing value at short
distances have strong empirical evidence. The proposed vortex picture is not QFT, and in view of SOC the coupling constant obtained from Eq.(108)
\begin{equation}
 \alpha_u = 1 + \frac{\alpha}{2\pi} -0.328478444 \frac{\alpha^2}{\pi^2}
\end{equation}
depends only on $\alpha$. How do we explain the important concept of the running coupling constant? Let us note that three vortex structure
having fixed coupling constants represents an isolated electron that we term as meta-electron; it may be seen analogous to a bare electron in QED.
The concept of meta-electron and the meaning of the coupling constant $\alpha_u$ given by Eq.(119) have to be understood clearly before we
proceed further. Meta-electron is a geometrical object, and the space-time vortices comprising it are the solutions of the massless
2-spinor wave equations. Fine structure constant $\alpha$ has a geometric origin, therefore, it is assumed to have  an invariant value. The unifying
coupling constant $\alpha_u$ has three terms that are interpreted to represent strong, electromagnetic and weak interactions. The manifestation
of any of these terms or that of the combined one depends on the nature of the physical environment in which the meta-electron
is embedded.

A simple example to iluustrate this idea is that of hydrogen atom. Hydrogen atom is electrically neutral though its internal structure has charged
constituents electron and proton that depending on the physical conditions of the experiment show plenty of physical phenomena. The observation
of hydrogen spectrum gave birth to the old quantum theory of Bohr, the later discoveries of fine structure and Lamb shift played
pivotal role in establishing QED. Formation of $H_2$ molecule also offers physical insights on the nature of effective interaction between neutral
hydrogen atoms in both Heitler-London method and molecular orbital approach.

Physical environment in the most common situations seem to be such that the observed electron shows only the presence of the electromagnetic
interaction that we associate with the second vortex in the expression (119); however the other two terms do not vanish, they
only remain in a dormant or latent state. The geometric interpretation of $\alpha$  for the meta-electron assigns it a fixed value independent of the 
physical environment. It is remarkable that the measured value of $\alpha$ from different kind of measurements yields a fixed value, see 
Table 2.3, section 2.12.3 in \cite{12}. Two vortex model of neutrino represents a meta-neutrino having the following coupling constant
\begin{equation}
 \alpha_\nu = 1  -0.328478444 \frac{\alpha^2}{\pi^2}
\end{equation}
This picture of meta-neutrino emerges from the idea that inside hadrons this object behaves like quarks. The question arises whether all the three
vortices independently or in pair-wise combination could also exist. From geometry point of view there is no argument against such a 
possibility. In fact, one may raise a question if higher-order QED contributions to the anomalous magnetic moment of electron in Eq.(107)
should also be not treated as possible geometric sructures similar to the three terms in (107). Geometrically this is allowed, however
the physics of vortices may restrict it to only three: the stability of vortex structures is crucial. A passive geometric spiral or any other
curve has a mathematical reality; the dynamical evolution of proposed geometrical structures here for meta-electron and meta-neutrino would
bring vortex-vortex interactions and physical environment to constrain the emerging stable configurations. We suggest that more than
three vortices may not be dynamically stable, and the three vortices for (119) and two vortices for (120) may be the only stable ones.

Note that SOC hypothesis is proposed to be applicable to only fundamental objects meta-electron and meta-neutrino in our model. Proton and 
neutron also have spin of $\hbar/2$ value, but we cannot apply SOC to them as they are composite structures. 
In the naive quark model a simple relation taking into account the spin of constituent quarks 
gives the spin of neutron and proton. In the present model, the knotted structures comprising meta-electron (positron) and meta-neutrino
determine the properties of neutron and proton.

Returning to the concept of running coupling constant, in the space-time vortex model the meta-electron has no electromagnetic field
associated with it in the conventional sense, but there could arise field like disturbances in the surrounding environment. Consistent
with the space-time picture a photon is envisaged as a composite vortex and the electromagnetic field is interpreted as a photon fluid \cite{28}.
Photon fluid and various kinds of vortices/vortex knots, collectively termed as physical environment, plays the role analogous to that of elementary
particles, field quanta and quantum vacuum of QFT and SM. Therefore it is natural to expect that the coupling constants in (119) would acquire
weight factors depending on energy scale and the physical environment. Statistical mechanics of vortices becomes important for the determination 
of the effective coupling constants. The formal analogy between QFT and statistical mechanics discussed in the literature, e. g. see
section 9.3 in \cite{39}, strengthens the idea that weight factors determine the effective coupling constants.  
There are two possible ways to introduce weight factors: for each term we have a separate weight factor or there is a single
weight factor  $\alpha \rightarrow W \alpha$. The effective coupling constant for (119) becomes
\begin{equation}
 \alpha_u^{eff1} = W_1 + W_2\frac{\alpha}{2\pi} - 30.328478444 W_3\frac{\alpha^2}{\pi^2}
\end{equation}
or
\begin{equation}
  \alpha_u^{eff} = 1 + \frac{W\alpha}{2\pi} -0.328478444 \frac{W^2\alpha^2}{\pi^2}
\end{equation}
In either case the negative sign in the last term acquires added significance since the effective coupling constant could become
vanishingly small at some value of energy dependent weight factor. Meta-neutrino coupling constant also becomes effective coupling constant, and 
assuming $\alpha \rightarrow W \alpha$ we have
\begin{equation}
  \alpha_\nu^{eff} = 1 -0.328478444 \frac{W^2\alpha^2}{\pi^2}
\end{equation}
A simple calculation shows that at $W \approx \alpha^{-1} \pi^{3/2}$ the effective coupling constant $\alpha^{eff}_\nu =0$. One may similarly
solve for $W$ setting $\alpha^{eff}_u =0$ in Eq.(122). Note that the vanishing QCD coupling constant obtained from RGE refers to only
strong interaction. In the present work Eq.(122) includes the effective interaction for all the three fundamental interactions; the limit
$\alpha^{eff}_u \rightarrow 0$ is markedly different than that in QCD $\alpha_s \rightarrow 0$.

The most important unsolved problem in modern theoretical physics is believed to be the unification of fundamental interactions in nature.
Why unification?  The ambitious unification program is not rooted in empirical or observable physics but it is inspired by pure thought: 
aesthetics and philosophical belief in the unity. SM of particle physics does have elements of unification, however it is founded on
a direct product of gauge groups having separate gauge couplings. Asymptotic freedom in QCD and the behavior of the electroweak coupling at
high energy motivated grand unified theories (GUT). GUT predicts unification of gauge couplings at energy scale of $\approx 10^{15}$Gev.
To get sensible meaning of gauge couplings it is recognized that $SU(2)\times U(1)$ and $SU(3)$ should be the sub-groups of a simple group,
GUT in Georgi-Glashow model is based on $SU(5)$ gauge group \cite{12}.
It is expected that at this energy scale gravity would become strong and unified theory must include gravity. Superstring theory sets this
goal of unification: a hypothetical 10D space-time becomes imperative \cite{18}. Past few decades have witnessed immense intellectual efforts
to develop superstrings. Unfortunately there is very little success in relating superstrings with the observed physical world of particles
and fields except the low energy limit where one gets gravity of general relativity. Physicists hope M-theory, twistors and other kind of
speculations in which space-time is emergent and/or illusion would bring next revolution in superstring paradigm to reach a final theory.
We differ on this; in fact, in our opinion \cite{15, 40} the main drawback of superstring paradigm is that of discarding space-time reality.
Reviewing twistor theory at 50 years \cite{41} the authors suggest holomorphic string theory in the twistor space as a promising
future in the unification goal. However, note that the twistor idea itself began discarding fundamental reality to space-time, and claiming
to deduce it from the objects like spin networks. Even this enthusiastic review \cite{41} admits very little impact of twistors on physics.

Postulated spin in twistors and 1D string replacing point in the space originally proposed in connection with the strong interaction, if given
space-time rendition, could become key ingredients in an alternative unification scheme. SOC and dynamical spiral and knots in our model
may be viewed as a significant effort in this direction; the present work is a step forward in a radically new approach to fundamental
questions in physics \cite{28}. The alternative unification paradigm has great virtues: it is internally consistent, it respects the 
cardinal principle in nature that of simplicity, parsimony and harmony, and restores the primacy to space-time reality. This claim is explained 
and elaborated in the following.

Unification has two aspects: search for the elementary constituents of matter, and unified theory of the fundamental interactions. In the 
contemporary scenario the first quest has landed in the unending sequence: matter $\rightarrow$ atoms $\rightarrow$ elementary particles $\rightarrow$
quarks $\rightarrow$ sub-quarks/preons $\rightarrow$ ? Instead of simplicity and parsimony the things have become increasingly complex
with a large number of the elementary constituents and their exotic properties. Compared to this, in the proposed model meta-electron (positron)
and meta-neutrino are the only elementary constituents, and these are also visualized in terms of space-time structures. However the space-time
picture is radically revised: it is not Newtonian with the imprint of Euclidean geometry, and it is not Minkowskian or pseudo-Riemannian 4D spacetime
of relativity.

A comprehensive critique on the conceptual foundations of space and time is presented in \cite{42}. In analogy to the fluid continuum, nontrivial
local spatial structures, topological defects/obstructions, and metric structure define the space continuum. Discreteness and one-to-one
correspondence with the natural numbers, and approximately the geometry of directed 1D line define time; see Chapter 4 in \cite{42}. To appreciate
the drastic revision envisaged here we mention that in conventional picture too one has Einstein-Rosen bridge, wormholes, and Goedel metric
representing typical nontrivial characteristics. It is also important to note that Einstein vacuum field equation could be viewed just a statement
on the geometries having vanishing Ricci curvature. Pseudo-Riemannian metric space-time has also been enriched postulating noncompact gauge group
of homothetic transformations in Weyl unified theory of gravitation and electromagnetism \cite{43}. The new proposition \cite{42} postulates
3D space and 1D time; the 4D spacetime metric in relativistic world-view is interpreted as 3D space having statistical fluctuations that
a point in space is not sharply defined \cite{42, 44}.

The unification of the fundamental interactions is embodied in the expression (119). It has perfect harmony in the sense that all the three
interactions have underlying unity represented by 2-spinor structure, and there is only single constant $\alpha$ comprising of the fundamental
constants $\hbar, e, c$. The magnitude of spin in the expression (108) defines the three coupling constants in the light of SOC. Does there
exist an analogue of the merging of coupling constants predicted in the unified electroweak theory and GUT? The concept of this kind of
unification does not exist in our model, however it is possible that under appropriate physical conditions the effective coupling constant 
defined by Eq.(122) has only a single effective interaction. For example, at the energy scale where $W \approx \frac{\pi^2}{2 \alpha}$ we have
$\alpha^{eff}_u \rightarrow 1$. In this limit the meta-electron has only strong interaction, and $\alpha^{eff}_\nu \rightarrow 1 - \pi/4$. Another
case is the energy scale where $\alpha^{eff}_\nu \rightarrow 0$, and $\alpha_u^{eff} \rightarrow \sqrt{\pi}/2$ that corresponds to only
effective electromagnetic interaction of the meta-electron.

In the present alternative scenario there is nothing like color of QCD gauge group, and the internal symmetries belonging
to the hypothetical spaces are redundant. Lepton number and flavor have certainly played important role in the classification of elementary
particles, however in our model there is no necessity of them. Space-time symmetries are the primary symmetries here, and there could
arise additional ``secondary symmetries'' associated with the combination of vortices to form various kinds of knotted structures.

Internal consistency of a theory, in our view, is decisive for it to represent physical reality: it must be free of logical
contradictions and must incorporate unambiguous experimental observations.  The proposed unification framework is logically consistent 
since no ad hoc or arbitrary assumption is made, and the flow of arguments is to a large extent akin to mathematical logic. The space-time
geometry has been given a radically new interpretation in \cite{42} analyzing the foundations of Newtonian space and time, and relativistic
4D spacetime continuum. The simplest nontrivial geometric structures in this new 3D space and 1D time space-time geometry are proposed to be
meta-electron and meta-neutrino. QED calculated magnetic moment of electron is assumed to have fundamental physical significance. The 
consequences of SOC hypothesis serve the basis of unified interactions. Complex combinations of meta-electron and meta-neutrino
forming knots are suggested to represent the observed elementary particles. There is a vast variety of knots, for example, 165 different knots
tabulated by Tait \cite{33} have upto 10 crossings. Pure geometry is made physical assuming space-time as a fluid, and in analogy to
fluid dynamics vortex interpretation is given to the geometric strings forming braids/knots. The vortex knots imply that among geometric
knot structures only the stable ones correspond to the elementary particles.

Geometrically an interesting approach is to construct vortex metrics using the Kerr-Schild form of the metric tensor \cite{27}. We have used
bi-scalar field to derive cylindrical vortex metric \cite{27}. It has to be genralized to obtain the vortex metric using massless 2-spinor
wave equation. In this method, the solution of Einstein field equation is not the objective, rather the constructed metric tensor
could be used to calculate various curvature tensors if desired. Vortex-vortex interaction and large number of vortices 
have to be understood using the theory of fluid dynamics \cite{37} and 
statistical mechanics of vortices. The idea of effective coupling constant in Eqs. (122) and (123) makes use of a phenomenological
weight factor $W$ that should be derivable from statistical methods. To make the idea more convincing let us have a look on
the QED running coupling constant, for example, in the approximate form given by Equation (2.360) in \cite{12}
\begin{equation}
 \alpha(Q^2) = \frac{\alpha}{1-(\alpha/2\pi)~ ln(Q^2/m^2)}
\end{equation}
The replacement $\alpha \rightarrow W \alpha$ in Eqs. (122) and (123) shows that $W$ may be given an energy-dependent functional form
\begin{equation}
 W= \frac{1}{1-A ~ ln(E^2/E_0^2)}
\end{equation}
Here $E_0$ is some reference energy scale and $A$ is a constant parameter. Logarithmic variation in (125) in analogy to running QED coupling
seems necessary for the consistency with the observations, see Figure 7.10 in \cite{39}.

The present unification framework offers a qualitative picture. Since the geometry of knots, vortices and fluid dynamics, and statistical 
mechanics are well established, and we envisage their prominent role in the theory for quantitative calculations we hope that a rigorous theory
for the present framework could be developed. Typically the electron-positron annihilation to $2\gamma$, to $\mu^- \mu^+$, and to
hadrons; positronium and its decays, and neutral pion and its decay are the problems that may be re-visited in the knotted vortex picture
and effective coupling constants $\alpha_u^{eff}$ and $\alpha^{eff}_\nu$ in our theory as test cases.

\section{\bf Discussion}

In the preceding section an alternative paradigm for unifying plethora of observed elementary particles and the fundamental interactions
has been articulated. Most notable is the departure from the standard approach in not using QFT. Utility of QFT is well known \cite{12, 18, 35, 39}.
Witten makes it quite explicit \cite{45} asserting that, 'the framework of special relativity plus quantum mechanics is so rigid that it practically
forces quantum field theory on us'. Here Witten, in fact, implies orthodox or Copenhagen interpretation of quantum mechanics. However the orthodox
interpretation is not a final word, and many alternative interpretations continue to challenge it \cite{1, 21, 38}. Contrary to Witten's
view-point we have argued that unification needs alternative approach without QFT \cite{46}. Sections II, III, IV and V in the present paper are 
devoted towards this goal. The key problems in the foundations of quantum theory are identified to be the nature of the complex
wavefunction, the meaning of the imaginary unit $i$ in the wave equations, and the origin of spin. Real wave equations obtained using 
the transformation (6) and introducing topological obstructions in geometry and algebra lay the foundation for the proposition of SOC in
section VI. SOC unifies 'charge' and 'spin' rendering internal symmetry groups like U(1) and $SU(3)_c$ in SM superfluous. To put the 
present work on $i$ and real wave equations in perpective we discuss earlier contributions by Stueckelberg \cite{47}, Segal \cite{48}
besides extensive work by Hestenes \cite{24}.

Stueckelberg \cite{47} is concerned with the question ``why the imaginary unit enters quantum theory''. Quantum theory in real Hilbert
space is developed by him introducing an operator $\hat{J}$
\begin{equation}
 \hat{J}^2 = - 1
\end{equation}
This operator commutes with all observable operators and replaces $i$ in quantum mechanics in complex Hilbert space. Observables are 
symmetric tensors. The necessity for antisymmetric operator $\hat{J}$ follows from the considerations on the commutators
of observable operators and the Heisenberg uncertainty relations. Interestingly the C-matrix (5) is antisymmetric and the transformation (6)
in the wave equations is analogous to the introduction of $\hat{J}$ operator in the commutators in the operator formalism developed by Stueckelberg. 
Recalling the formal equivalence between matrix mechanics and wave mechanics in quantum theory proved by Schroedinger, p.127 \cite{16}, the 
real Hilbert space operator formalism \cite{47} may be viewed as complimentary approach to our work on real wave equations. However, one must
recognize the conceptual difference between wave mehanics and matrix mechanics that discards the space-time representation \cite{13}.
Schroedinger himself pointed out that matrix mechanics was a ``true discontinuum theory'' in contrast to his ``continuum theory''. The new 
insights that we obtained on relating $i$ with the spin state of the Schroedinger wavefunction, and also on the nature of $Z_2$ vortex
are not obvious in the formalism of \cite{47}. One of the important questions discussed in \cite{47} is on the ortho-chronous and pseudo-chronous
Lorentz transformations that deserves further exploration in the context of our approach.

Finkelstein \cite{49} underlines the role of what he calls 'the conceptual expansion' in the progress of physics and highlights Segal's paper \cite{48},
and also notes the limits of the concept of time both at very small and very large scales. Segal's insightful remarks on Lie algebra
corresponding to Heisenberg relations have been treated from physics point of view for the oscillator commutators by Finkelstein. The presence
of imaginary unit $i$ in the commutators and the assumption in quantum mechanics that all operators commute with it needs expansion. For example,
the operators $\hat{q} =i q ;~ \hat{p} = -i p$ satisfy
\begin{equation}
 \hat{q} \hat{p} - \hat{p} \hat{q} = \hbar i
\end{equation}
\begin{equation}
 i \hat{q} - \hat{q} i =0
\end{equation}
\begin{equation}
 i \hat{p} - \hat{p} i =0
\end{equation}
The suggested expansion following Segal is
\begin{equation}
 \hat{q} \hat{p} - \hat{p} \hat{q} = \hbar i
\end{equation}
\begin{equation}
  i \hat{q} - \hat{q} i =\hbar' \hat{p}
\end{equation}
\begin{equation}
 i \hat{p} - \hat{p} i =- \hbar'' \hat{q}
\end{equation}
Finkelstein terms these relations as stabilizing variations where $\hbar', \hbar'' > 0$ are Segal constants. Curiously the expanded commutators
(130)-(132) using the  rescaling
\begin{equation}
 \hat{q} = Q \hat{L}_1 ; ~ \hat{p} = P \hat{L}_2 ; ~ i=J \hat{L}_3
\end{equation}
lead to the angular momentum commutators
\begin{equation}
 \hat{L}_1\hat{L}_2 -\hat{L}_2\hat{L}_1 =\hat{L}_3
\end{equation}
\begin{equation}
 \hat{L}_2\hat{L}_3 -\hat{L}_3\hat{L}_2 =\hat{L}_1
\end{equation}
\begin{equation}
 \hat{L}_3\hat{L}_1 -\hat{L}_1\hat{L}_3 =\hat{L}_2
\end{equation}
where $J=\sqrt{\hbar' \hbar''} ; ~ Q= \sqrt{\hbar \hbar'} ; ~ P = \sqrt{\hbar \hbar''}$. The trans-quantum commutators (130)- (132) in which
$i$ does not commute with $\hat{q}, \hat{p}$ describe rotator rather than oscillator. The significance of this result in connection with the 
interpretation of the Schroedinger particle in spin eigenstate \cite{5, 17} and the present work is important since $i$ plays key role in
Segal-Finkelstein arguments on angular momentum. Segal \cite{48} discusses the importance of a 
scale factor on the operator group algebra, in particular, a non-Abelian
Lie algebra corresponding to the Heisenberg relations. Though Segal's emphasis is on abstract Lie algebra, geometry and partial differential
operators could bring his approach closer to the present work on complex to real wave equations.

The role of the imaginary unit $i$ in quantum theory stimulated the work by Stueckelberg \cite{47} and Segal \cite{48}. Does it unveil the mystery
associated with $i$ \cite{1} ? It seems these attempts have not been very successful. Hestenes has spent many decades on the theme of geometric
algebra and in the process developed an alternative interpretation called real quantum mechanics in which the imaginary unit is interpreted
as a bivector. Though we made a brief comment on this in section V, a detailed discussion would be useful to delineate the strength and weaknesses
of GA approach. It is significant that philosophical perception leads Hestenes to put forward two ideas: 1) One can create a unified mathematical
language for physics, and 2) a major task for theorists is to construct a mathematical language that optimizes expression of key ideas and 
consequences of the theory. According to him GA has offered new insights into the structure of physics, and it provides a unified
language for whole of physics. We may remark that there is no new result or prediction in physics using GA. Why?

New insights into the structure of physics is a passive contribution having hardly any scope for a creative pathway to fundamental questions in physics.
Relation between physics and mathematics has differing view-points though most physicists tend to consider mathematics as a language of physics.
In spite of this the claimed unified mathematical language for physics would appear most ambitious and highly questionable. Utility
of known mathematics as a tool for physical theory, and the invention of new mathematics in the light of experiments and observed phenomena
have found many applications in the development of physics. Insightful remarks on geometry-physics-mathematics could be found at various places
in \cite{50}. Specifically the articles by Chern, Chapter 16; Regge, Chapter 17; discussion remarks on page 284; and Yang's views in Penal 
Discussion, Part XII \cite{50} are quite illuminating. We believe that language by its very nature has intrinsic limit to represent truth in totality,
specially in the domain of subtle and abstract reality. Note that language is basically a tool or mode of the expression of something, for example,
our thoughts or feelings. If mathematics is a language of physics, the profound idea of mathematical truth(s) that may not be in tangible form,
are likely to get defiled. Contrary to this the representation of partial truths in manifest physical phenomena would make physics to be a
natural language of mathematics. In fact, we already do it in practice as physicists employing the art of approximations and error analysis
in theories. Basic concepts, for example, electric or magnetic field too depend on limiting process. In a comprehensive work \cite{42}
we have proposed that mathematical truth(s) are the nearest intangible manifestations of reality, and physical phenomena represent their tangible
forms. This may be stated \cite{51} in the form of a hypothesis: Physics is the natural language of mathematics.

In the light of a brief commentary on the question of mathematical reality and the meaning of physics-mathematics relationship it could
be stated that there is a basic flaw at the level of motivation for developing GA. A short review on GA beginning with the work of Grassman
and Clifford is presented in section VIII of \cite{24}. Regarding the lack of desired reception of GA in the mainstream the past experiences
on matrix algebra in quantum theory and pseudo-Euclidean geometry in special relativity \cite{52}, and Regge's lament \cite{50} on ``the 
almost total neglect of the language of forms'' would bring a sobering impact. Minkowski adopted Euclidean 4D metric for spacetime assuming
imaginary time coordinate; and it was Born who recognized that Heisenberg relation was a noncommutative matrix multiplication law \cite{52}.

Geometry and quantum theory with reference to GA need further deliberation. Synthetic geometry, coordinate geometry, complex variables,
quaternions, vactor analysis, matrix algebra, spinors, tensors and differential forms are different mathematical systems but they have
a ``common geometric nexus'' \cite{24}. The claim that they constitute a highly redundant system \cite{24} is incorrect: any postulated
core geometric concept that supposedly unifies this system and eliminates redundancy would necessarily be postulate-specific. Beautiful
diverse aspects may get wiped out in this process. Moreover the importance of topology for global description of physical phenomena cannot
be ignored in any geometric framework. Instead of coordinate geometry and tensors Cartan's method of differential forms captures the
essence of metric-independent topologiocal properties in a nice way \cite{38}. This, of course, does not mean that metric property is redundant
or useless; majority of physics experiments pertain to local data collection, for example, in particle physics one measures cross sections.
In Cartan theory of spinors \cite{3} one finds judicious synthesis of metric space, spinor and matrix algebra. It is also remarkable
that complex wavefunction was crucial for the discovery of Aharonov-Bohm effect and Berry phase that provided impetus to the geometry
of quantum state space \cite{53}. We think that the mathematical systems listed above are complementary to each other and one may
seek connecting threads between them. To give an example, we have arrived at a new result in section II connecting C-matrix and $i$ 
introducing a topological point defect: it brings out the spin state of a Schroedinger particle in a clear way.

Logical extensions may also prove useful: a notable example is that of the metric geometry of complex Hilbert space \cite{52}. Another good
example is Maxwell equation: representation in terms of (${\bf E},{\bf B}$) vectors, covariant form using the electromagnetic field tensor
$F^{\mu\nu}$, and in differential form. In such cases one has to keep in mind the fact that though extended secondary constructs have
utility in specific problems they are not fundamental \cite{15, 42, 54}. If one reverses the logic treating them fundamental for building
a framework to the primordial, one is bound to end up in artificiality and superfluousity. Space and time as emergent or constructs based on twistors
\cite{41}; from Hilbert space and entanglement, and speculations like ER=EPR \cite{55}; and superstrings have not succeeded for 
the simple reason that space and time are intuitively perceived objects closer to reality \cite{15, 40, 42}.

To better appreciate GA approach to quantum mechanics we refer to a nice exposition on multivector algebra \cite{56} in addition to \cite{24}.
In real quantum mechanics proposed by Hestenes the imaginary unit and Pauli spin matrices are re-interpreted in GA version. Briefly stated the 
geometric product of vectors ${\bf a}, {\bf b}$ in a real vector space has a canonical decomposition
\begin{equation}
 {\bf ab} = {\bf a}.{\bf b} + {\bf a}\wedge {\bf b}
\end{equation}
where symmetric inner product ${\bf a}.{\bf b} = \frac{1}{2} ( {\bf a b} +{\bf b a})$ and antisymmetric outer product 
${\bf a} \wedge {\bf b} =\frac{1}{2} ( {\bf a b}- {\bf b a}) = -{\bf b} \wedge {\bf a}$; the former is scalar and the later is called a bivector
that geometrically represents an oriented plane segment. Bivector is intrinsic to the plane containing ${\bf a},{\bf b}$ and differs
from the standard vector product ${\bf a}\times{\bf b}$ in this sense. To develop GA an orthonormal set of vectors ${\bf e}_i;~i=1.2....N$
is defined for N-dimensional Euclidean space
\begin{equation}
 {\bf e}_i.{\bf e}_j =\delta_{ij}
\end{equation}
In 2D a unit bivector
\begin{equation}
 {\bf i} = {\bf e}_1{\bf e}_2={\bf e}_1 \wedge {\bf e}_2 =-{\bf e}_2{\bf e}_1
\end{equation}
has the property that
\begin{equation}
 {\bf i}^2 = -1
\end{equation}
Thus geometrically $\sqrt{-1}$ is identified as a unit bivector ${\bf i}$. Operationally ${\bf e}_2 ={\bf e}_1 {\bf i},~{\bf e}_1 = {\bf i}{\bf e}_2$.
Unit bivector ${\bf i}$ represents a unique oriented area for the plane, and rotates the vectors in the plane through normal angle.

Two unit vectors ${\bf u},{\bf v}$ having angle $\theta$ from ${\bf u}$ to ${\bf v}$ define a rotor
\begin{equation}
 U = {\bf u v} = \cos \theta + {\bf i} \sin \theta = e^{{\bf i} \theta}
\end{equation}
Using the rotor one can interpret the geometric product of arbitrary vectors ${\bf a b }$ as a complex number $z= \lambda U ={\bf a b}$. 
Geometric interpretation of $U$ is directed arc on a unit circle. GA offers a simplified calculational tool for complex variables.  On the 
other hand, Abelian unitary group U(1) defined by Eq.(11) and in the equivalent form (12) using the C-matrix transformation (6) along
with the geometrical interpretation of a directed 1D line has led us to discover the topological feature associated with the 
imaginary unit in section III-B. Topological origin of spin and Schroedinger particle in spin state are its important implications.
It may be repeated that the suggestion in \cite{5} regarding Schroedinger wavefunction to be that given by Eq.(54) is independent of GA.

In 3D Euclidean space the set of orthonormal vectors ${\bf e}_1,{\bf e}_2,{\bf e}_3$ define a unique trivector
\begin{equation}
 i={\bf e}_1{\bf e}_2{\bf e}_3
\end{equation}
and a bivector basis
\begin{equation}
 {\bf e}_1{\bf e}_2= i {\bf e}_3;~{\bf e}_2{\bf e}_3= i {\bf e}_1; ~{\bf e}_3{\bf e}_1= i {\bf e}_2
\end{equation}
Here $i$ is a pseudo-scalar and has the property that of the imaginary unit
\begin{equation}
 i^2 =-1
\end{equation}
An important relation follows using the multiplication by $i$ relating the outer product with the conventional vector product 
${\bf a} \wedge {\bf b} = i {\bf a} \times {\bf b}$. Geometrically the pseudo-scalar $i$ is interpreted as an oriented unit volume, and 
unit bivectors (143) a basis of directed areas in planes with orthogonal intersections. One can easily verify that the relations (143) are 
similar to that satisfied by Pauli matrices: replace $e_i \rightarrow \sigma_i$. In \cite{24} the symbol $\sigma_i$ is used to make the similarity
obvious, however it is better to keep the distinct notation since the conclusion that Pauli algebra is 
a matrix representation of GA needs arguments given in  \cite{24}.

Spin and 2-spinor Pauli equation are re-visited in GA \cite{24}. Imaginary unit in Pauli matrices (14) is denoted by $i'$ termed scalar
imaginary having no geometrical or physical significance. Real GA version of Pauli matrix theory is obtained using $\Psi_{Pauli} \rightarrow \psi u$
where $u$ is the eigenvector of $\sigma_3$ with eigenvalue +1 in (97). Using Pauli algebra $\sigma_1 \sigma_2 u = i' u$. It is then shown that
$\psi$ is a polynomial in $\sigma_i$ with real coefficients. Now using the correspondence $\sigma_i \rightarrow e_i$ one has 2-spinor
$\Psi_{Pauli} \rightarrow ~$ even multivector $\psi$ in GA. Even multivector $\psi$ has a scalar part and three bivectors, and it 
is interpreted as real spinor in GA. 

Subsequent discussion \cite{24} is just a re-interpretation of the standard quantum mechanics with no new result. However the Pauli algebra
(105) and its physical interpretation in accordance with the Heisenberg relation is rejected in the GA interpretation. The argument is that
the commutation relation
\begin{equation}
 [ e_1, ~e_2] =2 e_1 \wedge e_2
\end{equation}
with the correspondence $e_1 \rightarrow \sigma_1, ~ e_2 \rightarrow \sigma_2$ would be just a geometric product. In this connection, one of the 
important new results obtained by us deserves mention: departing from both standard quantum mechanics and its GA version we have uncovered 2D
topological obstruction, i. e. area discontinuity following the relation (105). This establishes the topological origin of spin. Could
geometrically directed area interpretation for the bivectors (143) in GA \cite{24}  be viewed as hiding the present topological interpretation?
In fact, one could speculate that the geometric interpretation of the trivector or pseudo-scalar (142) as oriented volume may have topological and/or
physical significance.

Spin-charge relation in the present paper is unconventional: it is radically different than original Weyl gauge theory \cite{43} as well as
modern gauge field theories \cite{12}. We have explained in section VI that this idea is not unphysical. We approach this question from other angle
that relates with Infeld-van der Waerden formalism. Spinors in curved spacetime were treated in this work in 1933 using 2-spinor formalism 
of van der Waerden \cite{19}. A nice review \cite{57} discusses this formalism. Spinor analysis in analogy to the tensor analysis
in the pseudo-Riemannian spacetime of general relativity is developed in section X of \cite{57}. The basic object is a metric spinor;
generalization of C-matrix (5) or Levi-Civita symbol in \cite{19}
\begin{equation}
 \gamma _{lm} = C \gamma_{12}
\end{equation}
Here $l,m$ indices take values $1,2 (\dot{1} \dot{2})$. Metric spinor could be expressed in terms of a complex number; setting 
$\gamma = \gamma_{12} \gamma_{{\dot{1}} {\dot{2}}}$ one may represent (146) as
\begin{equation}
 \gamma_{lm} = C \sqrt{\gamma} e^{i\theta}
\end{equation}
Spinor affinity, $\Gamma^l_{~m \mu}$ similar to Christofell symbol is obtained defining covariant derivative of spinors in world spacetime coordinates. 
Gauge covariant Dirac equation is also discussed in this formalism. An important result is obtained that the quantity 
$\Gamma^l_{~l\mu} -\Gamma^{\dot{l}}_{~{\dot{l}} \mu}$ transforms exactly like a 4-vector that occurs in Weyl gauge theory \cite{43}. Physical 
interpretation of the 4-vector gauge potential in Weyl geometry depends on the identification of the distance curvature, a geometric quantity, 
with the physical electromagnetic field tensor. The 4-vector gauge potential in Infeld-van der Waerden formalism is also interpreted as
electromagnetic potential. However, unlike vector length or scale change in Weyl theory here the spinor phase generates the 4-vector gauge potential. 
This amounts to a spin-charge relation of a particle. Penrose states that this theory contradicts the observed fact since neutron has 
spin but it is electrically neutral \cite{58}. This objection is not valid if one restricts the applicability of spin-charge relation to
the particles that have intrinsic spin not the composite one. In SM and QCD neutron has spin-half constituents, namely the quarks which have fractional
charge. Neutrino would be inconsistent with spin-charge relation. In our model, meta-neutrino has strong and weak charges that we have related 
with spin. In this generalized sense based on SOC there is no conflict with spin-charge relation. Would it be possible to reformulate
Infeld-van der Waerden formalism for real spinors?

\section{\bf Conclusion}

An insightful historical journey of SM presented by Weinberg \cite{59} ends with the question: What next? In the present paper a radically
new outlook on particle physics is suggested in which SOC hypothesis plays the key role. Spin and charge are two facets of the same
underlying reality and the differing magnitude of spin angular momentum determines the coupling strengths of strong, electromagnetic,
and weak interactions are the main new contributions of the present work. The hypothetical internal spaces for gauge symmetries in SM
become unnecessary since spin relates with space-time symmetry. Topological significance of the imaginary unit $i$; rendition of
complex wave equations to real ones using the transformation $i$ to C-matrix, Eq.(6); and physical interpretation of $\hbar$ in the 
wave equations as a unit of angular momentum that could be replaced by another unit constitute the main steps to establish the spin origin
of charge hypothesis.

Continuous spin, zero-mass wave equations irreducible under Poincare group \cite{2} show that double-valuedness of spinor wavefunction and 
arbitrary continuous magnitude of SAM are compatible, Eq.(16). Topological origin of spin explains discrete SAM values. Point defect in 1D 
directed line is proposed to give topological meaning to $i$, and a directed/oriented 2D area element as a topological obstruction is suggested
to be hidden in the commutation relations of Pauli matrices, Eq.(105). These propositions throw new light on $Z_2$ vortex \cite{4} and using
the recent work \cite{26} on angular momentum of pure gauge potential in the Aharonov-Bohm effect spin origin to topology arises once again:
topological origin of spin is established from various arguments. The detailed work on real wave equations is put on perspective discussing
the past literature on real quantum mechanics in GA \cite{24}, real Hilbert space quantum theory \cite{47}, and Segal's extended
commutator theory treating $i$ as an operator \cite{48}.

On particle physics and unification we have set a modest goal: developing a conceptual framework. Implication of SOC hypothesis on
the elementary constituents of matter leads to the concept of meta-electron and meta-neutrino as the only fundamental objects: nontrivial
geometry and topology of space-time, vortex knots, and the stability constraint are the main ingredients of particle model. Unification of 
interactions is discussed in terms of the effective coupling constant obtained by introducing the weight factors in the meta-electron
and meta-neutrino coupling constants.

To conclude, a synthesis of vortex dynamics, geometry and topology
of knots, and statistical mechanics of vortices is envisaged as a viable theory to be developed. We hope that the new ideas would stimulate
further work in this direction.

{\bf Acknowledgment}

I thank A. Afgoustidis, France for reference 48.

\end{document}